\documentclass[12pt]{article}
\usepackage{amssymb,slashed}
\usepackage{amsmath,euscript}
\usepackage{graphicx}
\usepackage{color}

\usepackage{hyperref}

\newcommand{\bmat}{\left(\begin{array}}
\newcommand{\emat}{\end{array}\right)}

\def\yzero{\smash{\hbox{$y\kern-4pt\raise1pt\hbox{${}^\circ$}$}}}

\def\beq{\begin{equation}}
\def\eeq{\end{equation}}
\def\beqa{\begin{eqnarray}}
\def\eeqa{\end{eqnarray}}

\def\-{\hphantom{-}}
\def\ov{\overline}
\def\s2{\frac{1}{\sqrt2}}

\def\beq{\begin{equation}}
\def\eeq{\end{equation}}
\def\beqa{\begin{eqnarray}}
\def\eeqa{\end{eqnarray}}

\def\Tr{{\rm Tr \,}}

\def\IF{\relax{\rm I\kern-.18em F}}
\def\II{\relax{\rm I\kern-.18em I}}

\def\Dsl{\,\raise.15ex\hbox{/}\mkern-13.5mu D} 

\def\IC{\bf C}
\def\IS{\bf S}
\def\IR{\bf R}
\def\IZ{\bf Z}
\def\IT{\bf T}
\def\IP{\bf P}
\def\NN{{\cal N}}

\def\aD3{${\ov{{\rm D}3}}$}

\catcode`\@=11   
\newdimen\@rotdimen
\newbox\@rotbox  

\def\@vspec#1{\special{ps:#1}}
\def\@rotstart#1{\@vspec{gsave currentpoint currentpoint translate
   #1 neg exch neg exch translate}}
\def\@rotfinish{\@vspec{currentpoint grestore moveto}}
\def\@rotr#1{\@rotdimen=\ht#1\advance\@rotdimen by\dp#1%
   \hbox to\@rotdimen{\hskip\ht#1\vbox to\wd#1{\@rotstart{90 rotate}%
   \box#1\vss}\hss}\@rotfinish}
\def\@rotl#1{\@rotdimen=\ht#1\advance\@rotdimen by\dp#1%
   \hbox to\@rotdimen{\vbox to\wd#1{\vskip\wd#1\@rotstart{270 rotate}%
   \box#1\vss}\hss}\@rotfinish}%
\def\@rotu#1{\@rotdimen=\ht#1\advance\@rotdimen by\dp#1%
   \hbox to\wd#1{\hskip\wd#1\vbox to\@rotdimen{\vskip\@rotdimen
   \@rotstart{-1 dup scale}\box#1\vss}\hss}\@rotfinish}%
\def\@rotf#1{\hbox to\wd#1{\hskip\wd#1\@rotstart{-1 1 scale}%
   \box#1\hss}\@rotfinish}%
\def\rotate{\@ifnextchar[{\@rotate}{\@rotate[l]}}
\def\@rotate[#1]#2{\setbox\@rotbox=\hbox{#2}\@nameuse{@rot#1}\@rotbox}

\catcode`\@=12

\topmargin
-1.5cm
\textwidth
15.5cm
\textheight
23.5cm
\oddsidemargin
0.7cm
\evensidemargin
1.2cm

\begin{document}

\makeatletter
\@addtoreset{equation}{section}
\makeatother
\renewcommand{\theequation}{\thesection.\arabic{equation}}
\pagestyle{empty}
\rightline{ IFT-UAM/CSIC-09-21}
\rightline{ CERN-PH-TH/2009-045}
\vspace{0.1cm}
\begin{center}
\LARGE{\bf Non-perturbative effects and wall-crossing from topological strings  \\[12mm]}
\large{Andr\'es Collinucci$^\dagger$, Pablo Soler$^\ddagger$, Angel M. Uranga$^{\ddagger \, *}$\\[3mm]}
\footnotesize{$^\dagger$ Institute for Theoretical Physics, Vienna University of Technology\\
Wiedner Hauptstr. 8-10, 1040 Vienna, Austria}\\
\footnotesize{$^\ddagger$Instituto de F\'{\i}sica Te\'orica UAM/CSIC,\\[-0.3em]Universidad Aut\'onoma de Madrid C-XVI, Cantoblanco, 28049 Madrid, Spain} \\[2mm] 
\footnotesize{$^*$ PH-TH Division, CERN CH-1211 Geneva 23, Switzerland}\\

\small{\bf Abstract} \\[5mm]
\end{center}
\begin{center}
\begin{minipage}[h]{16.0cm}

We argue that the Gopakumar-Vafa interpretation of the topological string partition function can be used to compute and resum certain non-perturbative brane instanton effects of type II CY compactifications. In particular the topological string A-model encodes the non-perturbative corrections to the hypermultiplet moduli space metric from general D1/D(-1)-brane instantons in 4d $\NN=2$ IIB models. 
We also discuss the reduction to 4d $\NN=1$ by fluxes and/or orientifolds and/or D-branes, and the prospects to resum brane instanton contributions to non-perturbative superpotentials. We argue that the connection between non-perturbative effects and the topological string underlies the continuity of non-perturbative effects across lines of BPS stability. We also confirm this statement in mirror B-model matrix model examples, relating matrix model instantons to non-perturbative D-brane instantons. 
The computation of non-perturbative effects from the topological string requires  a 3d circle compactification and T-duality, relating effects from particles and instantons, reminiscent of that involved in the physical derivation of the Kontsevich-Soibelmann wall-crossing formula.

\end{minipage}
\end{center}
\newpage
\setcounter{page}{1}
\pagestyle{plain}
\renewcommand{\thefootnote}{\arabic{footnote}}
\setcounter{footnote}{0}


\tableofcontents


\section{Introduction}

Non-perturbative effects in string theory are a key ingredient in the proper understanding of the theory, and in particular of the dynamics of compactifications to four dimensions.  The formal developments on euclidean brane instanton effects (see e.g. \cite{Becker:1995kb, Witten:1996bn, Harvey:1999as, Witten:1999eg}), in particular D-brane instantons in type II compactifications (or F/M-theory duals) have led to a variety of (in some cases very explicit) applications to e.g. moduli stabilization \cite{Kachru:2003aw,Denef:2004dm,Denef:2005mm}, and generation of perturbatively forbidden couplings \cite{Blumenhagen:2006xt,Ibanez:2006da,Florea:2006si}. 

An important question for such non-perturbative effects is their behaviour under motion in moduli space, in particular in crossing walls of BPS stability  of the underlying D-brane instantons. The question was addressed in the 4d $\NN=1$ context in \cite{GarciaEtxebarria:2007zv} for non-perturbative superpotentials and in \cite{GarciaEtxebarria:2008pi} for multi-fermion F-terms. The physical picture is that wall-crossing of instantons is such that the resulting 4d non-perturbative terms are continuous across them, consistently with supersymmetry of the low energy effective action. A related discussion was carried out  in \cite{Gaiotto:2008cd}, for 4d $\NN=2$ supersymmetric field theories compactified on a circle, where continuity of non-perturbative corrections was shown to be equivalent to the wall-crossing formula in \cite{ks}.

These results can be rephrased as the statement that non-perturbative terms are insensitive to the stability conditions of the underlying BPS objects. More abstractly, 
they would be defined at the level of the category of holomorphic D-branes \cite{Douglas:2000gi}. This suggests a link, already hinted at in \cite{GarciaEtxebarria:2007zv}, with the topological string, whose D-branes are defined without any stability condition. It is then natural to imagine that the topological string captures certain D-brane instanton non-perturbative terms of the 4d physical theory. Thus, the insensitivity of the 4d non-perturbative terms with respect to walls of BPS stability is explained by the insensitivity of the underlying topological string theory, which computes them. 

In this paper we take some steps into fleshing out this proposal. We show that the Gopakumar-Vafa interpretation of the topological A-model on a CY $X$ computes D1/D(-1)-brane instanton effects for 4d $\NN=2$ type IIB theory on $X$. The argument involves the computation of a one-loop diagram of wrapped BPS brane particles, and application of the c-map, which effectively turns BPS particles into BPS D-brane instantons. In the D1/D(-1) brane instanton sector, there are walls of theshold stability in which a BPS instanton decays into several mutually BPS instantons. The continuity of the corresponding non-perturbative effect, which microscopically requires conspiracies involving multi-instanton processes, becomes manifest in terms of the topological string, and follows from the invariance of the index for quantum BPS particle states. 

The D1/D(-1) instanton sector also contains walls of marginal stability, in which the decay products are mutually non-local. In that case, the continuity of the corresponding non-perturbative effect is best understood in terms of the M-theory lift of the instanton, where such a decay becomes energetically unfavorable.

We explore related results in three additional directions. First we show that the above picture holds in certain topological string models admitting a non-perturbative definition in terms of matrix models. Secondly, we study the extension of the above picture to 4d $\NN=1$ models, where threshold stability constitutes the only wall crossing phenomenon for D-brane instantons contributing to the superpotential. Finally, we explore the prospects to include more general (and mutually non-local) D-brane charges in the topological string, by using the 4d-5d connection relating the GV invariants to the BPS index for D6/D2/D0-brane particles (or D5/D1/D(-1)-brane instantons). We propose that such relation holds in a suitable sense throughout moduli space. We also show that
the topological string includes the relevant non-perturbative effects at least in a certain linear approximation and for a suitable charge sector.

The paper is organized as follows. In Section \ref{review} we review some ideas on the behaviour of non-perturbative effects across walls of BPS stability of the underlying D-brane instantons, and provide some examples. In Section \ref{topological} we describe the connection between the topological string and certain non-perturbative D-brane instanton effects, via use of the c-map (Section \ref{from4dto3d}) on the GV interpretation of the topological string (Section \ref{gvargu}). The implications for threshold and marginal wall crossing are considered in Sections \ref{threshold-walls} and \ref{marginal-walls}. In Section \ref{3daction} we present an explicit computation of D1/D(-1)-brane instanton corrections to the type IIB hypermultiplet metric, to which the above general arguments apply. In Section \ref{npdefinitions} we consider non-perturbative effects and wall crossing in examples of topological string B-model defined in terms of matrix models, and propose a new physical interpretation for matrix model instantons. In Section \ref{nequal1} we describe mechanisms to reduce the supersymmetry to 4d $\NN=1$, and describe aspects of non-perturbative effects and threshold wall crossing in those setups. In Section \ref{general-charges} we describe the prospects to include more general D-brane charges using topological strings. We discuss the inclusion of D5/D1/D(-1)-brane instanton charges, and its implication for wall crossing of the BPS index of 4d D6/D2/D0-brane particles (generalized Donaldson-Thomas invariants). In Section \ref{conclusions} we offer some concluding remarks. 

During preparation of this revised version, the paper \cite{Aganagic:2009kf} appeared, which has some overlap with Section \ref{general-charges}. It would be interesting to relate both pictures of wall crossing for general charges.

\section{Review of 4d non-perturbative effects and walls of BPS stability}
\label{review}

\subsection{Marginal stability vs. threshold stability}

In this section we review the microscopic description of walls of BPS stability, and the known results about the behaviour of non-perturbative effects across them in 4d $\NN=1$ \cite{GarciaEtxebarria:2007zv,GarciaEtxebarria:2008pi}  and $\NN=2$ theories \cite{Gaiotto:2008cd}.

In type II compactifications (or orientifolds thereof) on CY threefolds, the spectrum of BPS D-branes can jump across (or at) real codimension one walls in moduli space.  In our setup these D-branes can correspond to 4d particles or 4d instantons, depending on the 4d spacetime structure of the D-brane. 
It is convenient to distinguish, following  \cite{deBoer:2008fk,deBoer:2008zn}, the concepts of threshold stability and of genuine marginal stability. 

\subsubsection*{Wall of marginal stability}

At a wall of marginal stability, a BPS D-brane on one side of the wall splits into two D-branes at the wall, which misalign their BPS phases on the other side of the wall, thus making the overall object non-BPS.

In the case of D-instantons, the real parameter $\xi$ parametrizing the direction transverse to the wall is a K\"ahler modulus in type IIB compactifications, and a complex structure modulus in type IIA models, and couples as a Fayet-Illiopoulos (FI) term to the D-brane worldvolume theory. This allows for a microscopic description of the D-brane system at the wall, where the BPS D-brane is split into components \cite{Kachru:1999vj}. For simplicity, consider a 4d $\NN=2$ compactification, so that the worldvolume theory on the BPS D-brane system is the dimensional reduction (to zero or one dimensions, for D-brane particles or instantons) of a 4d theory with four supercharges. For 4d $\NN=1$ type II orientifolds, suitable orientifold projections of the forthcoming worldvolume descriptions would lead to similar wall structures.

Consider for concreteness a split into two  components. A typical wall of marginal stability is described by a worldvolume theory with gauge group $U(1)$ (the relative $U(1)$ of the two branes) and a charged chiral multiplet $\phi$, with charge normalized to $+1$, with no superpotential, and with D-term potential
\beqa
V_D\, =\, (\, |\phi|^2\, +\, \xi\,)^2
\eeqa
Hence for $\xi=0$, we have $\phi=0$, and the relative $U(1)$ and supersymmetry are unbroken, corresponding to a BPS system of two D-branes. For $\xi<0$, a vev for $\phi$ restores supersymmetry but breaks the $U(1)$, corresponding to a single bound BPS D-brane. For $\xi>0$, there is a non-supersymmetric minimum at $\phi=0$, with unbroken $U(1)$, corresponding to a non-BPS system of two D-branes. There is a manifest discontinuity in the spectrum of BPS D-branes in the system. Similar walls exist with additional number of chiral multiplets, as long as all carry equal sign charges. Hence a condition to have marginal stability walls is that sub-objects have non-zero intersection number (or DSZ product), which counts the (net) multiplicity of such fields.

\subsubsection*{Wall of threshold stability}
At a wall of threshold stability, the BPS D-brane splits at the wall, but the decay products recombine back into a BPS state at the other side of the wall.

A typical wall of threshold stability is described by a theory with gauge group $U(1)$ and two chiral multiplets $\phi_1$, $\phi_2$ with opposite charges $\pm 1$. Hence the D-term potential reads
\beqa \label{dtermthreshold}
V_D\, =\, (\, |\phi_1|^2\,-\, |\phi_2|^2\, +\, \xi\,)^2
\eeqa
There may potentially exist a superpotential or not, depending on the model, but we need not consider either possibility at the moment.
For $\xi=0$ we have a BPS system of two D-branes, while for $\xi\neq 0$ the D-brane forms a single BPS bound state, regardless of the sign of $\xi$. The spectrum of BPS D-branes is potentially changed only at the wall (but remains unchanged away from it).
Hence typically walls of threshold stability arise when the sub-objects have zero intersection number (or DSZ product), leading to zero net chirality for the modes of open strings stretched between them.

\subsection{Non-perturbative effects across walls} 
\label{nonpertacrosswalls}

Consider the computation of non-perturbative effects from D-brane instantons in a string compactification. The statement that non-perturbative F-terms arise from BPS D-brane instantons may suggest that they are discontinuous at real codimension one walls in moduli space, in plain contradiction with supersymmetry of the 4d effective action, which requires nice holomorphic dependence on the moduli. 

This puzzle was raised in  \cite{GarciaEtxebarria:2007zv} and addressed in several examples \cite{GarciaEtxebarria:2007zv,GarciaEtxebarria:2008pi} for 4d $\NN=1$ theories (see also \cite{Cvetic:2008ws}). The main results for D-brane instantons that contribute to the non-perturbative superpotential are

\begin{itemize}
\item BPS D-brane instantons that contribute to the superpotential must have exactly two fermion zero modes. Thus they cannot become non-BPS, since they do not have enough fermion zero modes for the four required Goldstinos. Hence such instantons
cannot cross walls of genuine marginal stability.

\item BPS D-brane instantons that contribute to the superpotential can cross walls of threshold stability. The decay products at the wall conspire to produce a superpotential contribution in a multi-instanton process, restoring holomorphic dependence of the superpotential on the moduli (and rendering it essentially independent of $\xi$).

\end{itemize}

BPS D-brane instantons not contributing to the superpotential, but to higher F-terms, can cross walls of marginal stability and become non-BPS. The non-perturbative amplitude is in a non-trivial class of the Beasley-Witten cohomology \cite{Beasley:2004ys,Beasley:2005iu}, so that locally they can be written as D-terms, but not globally due to an obstruction localized on the BPS locus. 

A prominent example of higher F-term correction in the $\NN=1$ setup is provided by corrections to moduli space metrics in $\NN=2$ theories. A beautiful systematic understanding of marginal wall crossing for BPS particles in 4d quantum field theories was provided in \cite{Gaiotto:2008cd} in terms of continuity of non-perturbative effects of BPS instantons in their 3d compactification. The result provides a physical interpretation of the wall crossing formula in \cite{ks}. A generalization for string compactifications is expected to hold, despite technical difficulties in making it completely precise.

In this paper we will be mainly interested in threshold wall crossing for D-brane instantons in 4d $\NN=2,1$ theories. For D-brane {\em particles}, such wall crossing are essentially harmless, as is well-known that the index counting BPS state degeneracies is continuous. As we discuss in Section \ref{threshold-walls}, even when the {\em classical} BPS D-brane particle splits, there is a {\em quantum} BPS bound state at threshold which keeps the index unchanged. For D-brane {\em instantons}, the splitting at the threshold wall implies the real disappearance of the single D-brane instanton contribution to the 4d non-perturbative effective action. Restoration of the continuity requires the existence of microscopically non-trivial multi-instanton processes. Our aim is to gain a deeper understanding of the appearance of these processes, both in $\NN=2$ and in $\NN=1$ theories (where they are particularly important, as the only wall crossing phenomenon for non-perturbative superpotentials).

\subsubsection*{Example of threshold walls for D1/D(-1)-brane instantons}

For future convenience it will be useful to introduce some explicit examples of threshold for D1/D(-1)-brane instantons. It is enough to consider examples of local CY's with compact 2-cycles. A large set of examples which have not appeared in the D-brane instanton literature, can be engineered the hypersurfaces $X_N$ in $\IC^4$ defined by
\beqa
xy\, =\, (z+w)\, (z+\alpha w) \ldots (z+\alpha^{N-1} w)
\eeqa
with $\alpha=e^{2\pi i/N}$. The model describes a complex surface fibration over a complex $w$-plane that degenerates to an $A_N$ singularity $xy=z^N$ over the origin $w=0$. Thus it contains $N$ compact 2-cycles $C_k$, $k=1,\ldots, N$ collapsed to zero size, with a homology relation $C_1+\ldots C_N=0$.
We are interested in D1-brane instantons wrapped on such genus zero curves $C_i$, with induced D(-1)-brane charge due to non-vanishing NSNS 2-form fields. Their BPS phases are controlled by the blowing up modes of the singularity, which are Kahler moduli and have walls of threshold stability in the blown-down limit. Under a T-duality in the $U(1)$ orbit $x\to e^{i\theta} x$, $y\to e^{-i\theta} y$, the geometries map to Hanany-Witten configurations of NS5-branes at angles (i.e. spanning the complex planes $z+\alpha^k=0$), with euclidean D0-branes suspended between them. Blowing up the singularity corresponds to the removal of a corresponding NS5-brane, and the recombination of the D0-brane segments. This construction allows an easy derivation of the spectrum and interactions on the instanton worldvolume. Essentially they correspond to dimensional reductions of affine $A_N$ quiver theories.

\begin{figure}
\begin{center}
  \includegraphics[width=7cm]{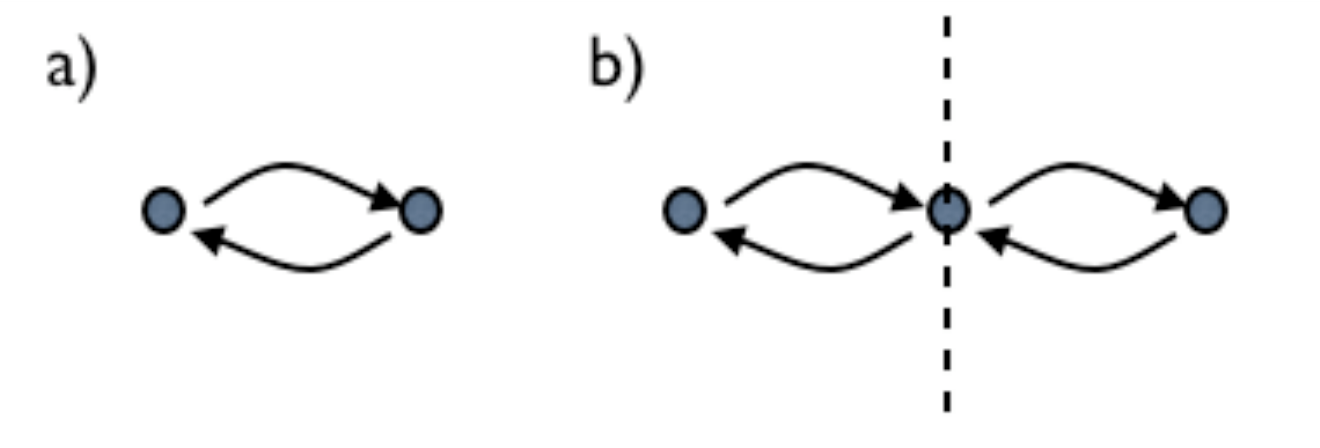}
  \caption{\small Typical quiver theories describing D-branes at  walls of threshold stability in 4d  $\NN=2$ (a) and $\NN=1$ (b) theories.}
  \label{quivers}
\end{center}
\end{figure}

Consider the geometry $X_3$ with D1-brane instantons on the 2-cycles $C_1$ and $C_2$. The blown-down limit corresponds to a threshold stability wall against recombination to a single D1-brane on $C_1+C_2$. The latter is a D1-brane instanton on a genus 0 curve, and has four fermion zero modes, thus leads to a correction to the hypermultiplet metric. In order for this to be continuous, there must be a 2-instanton process involving the D1-branes on $C_1$ and $C_2$ simultaneously. The latter system is described by a 0-dimensional supersymmetric quiver theory with four supercharges, obtained by dimensional reduction of a 4d $\NN=1$ $U(1)\times U(1)$ gauge theory, with two chiral multiplets $\Phi_{12}$, $\Phi_{21}$ with charges $(1,-1)$ and $(-1,1)$ respectively,  see Figure \ref{quivers}a. There is also a superpotential
\beqa
W\, =\, \Phi_{12} \Phi_{21}\Phi_{12}\Phi_{21}
\eeqa
Explicitly, the D1-brane instanton quiver action is
\beqa
  S & = & (x_1^\mu - x_2^\mu)^2\, (|\varphi_{12}|^2 +
  |\varphi_{21}|^2)\, +\,i(x_1^\mu - x_2^\mu)\, \{\, {\ov \chi}_{12} \sigma_\mu \chi_{12} -
  {\ov \chi}_{21} \sigma_\mu \chi_{21}\,\} \, + \nonumber \\
  & + &  (\chi_{12}\,(\theta_1 - \theta_2))\varphi_{12}^* -
  (\chi_{21}\,(\theta_1 - \theta_2))\varphi_{21}^* + ({\ov
    \chi}_{12}\,(\tilde\theta_2-\tilde\theta_1)\varphi_{12} - ({\ov
    \chi}_{21}\,(\tilde\theta_2-\tilde\theta_1))\varphi_{21}  \nonumber \\
&+&     \chi_{12}\varphi_{21}\chi_{12}\varphi_{21} \, +\, 
  2\chi_{12}\chi_{21}\varphi_{12}\varphi_{21} \, +\,
  \varphi_{12}\chi_{21}\varphi_{12}\chi_{21} \quad + \text{h.c.} \nonumber \\
&+&    (\, |\varphi_{12}|^2 - |\varphi_{21}|^2\, )^2 \, +\,  |\varphi_{21}\varphi_{12}\varphi_{21}|^2\, +\, |\varphi_{12} \varphi_{21}\varphi_{21}|^2
\label{two-instanton-zero-modes}
\eeqa
The first term describes the mass terms of the modes between the two instantons when they are separated in the 4d space. The second line describes the couplings of the difference between the two sets of Goldstinos $\theta_i$, $\tilde\theta_i$ on the two instantons. The third line contains the F-term fermion interactions. The last line describes  the (D-term plus F-term) potential for the bosonic modes. It is a tedious but straightforward exercise to check that configurations of coincident instantons (i.e. by localization onto $x_1=x_2$) can saturate all fermion zero modes except for the center of mass Goldstinos $\theta_1+\theta_2$, $\tilde\theta_1+\tilde\theta_2$. Hence the 2-instanton system has four overall exact fermion zero modes and provides the required non-perturbative contribution at the threshold stability wall.

In the 4d $\NN=1$ context one can obtain similar examples, by considering the geometry $X_4$ modded out by an orientifold action $\Omega R (-1)^{F_L}$ with $R:w\to -w$. The quiver theory is obtained by quotienting the affine $A_4$ quiver by the outer automorphism identifying the nodes $C_1\leftrightarrow C_3$ and leaving $C_2$ and $C_4$ invariant. Consider wrapping one D1-brane instanton on $C_2$, and on $C_1$ and its orientifold image $C_3$. The quiver theory, see Figure \ref{quivers}b is a 0-dimensional $SO(1)\times U(1)$ gauge theory, whose matter content and interaction are essentially the same as in the above $X_3$ model by simply removing the modes $\tilde\theta_2$. Using the interactions (\ref{two-instanton-zero-modes}) with $\tilde\theta_2=0$, it is straightforward to check that configurations of coincident instantons can saturate all fermion zero modes except for overall Goldstinos $\theta_1+\theta_2$. The 2-instanton system has two exact fermion zero modes and provides the required non-perturbative superpotential at the threshold stability wall.

\section{Non-perturbative effects from topological strings}
\label{topological}

In this Section we show that the Gopakumar-Vafa (GV) interpretation of the topological A-model on a CY $X$, upon compactification to 3d and T-duality,  computes non-perturbative D1/D(-1)-instanton effects in 4d $\NN=2$ type IIB on $X$.
Although the different steps in the argument are not new, the logic running through them is new, and leads to a useful computational tool for non-perturbative D-brane instantons effects, which we exploit in an explicit computation in Section \ref{3daction}.
In Sections \ref{threshold-walls}, \ref{marginal-walls} we argue that the continuity of the GV index for BPS particles underlies and explains the continuity of the non-perturbative D-brane instanton effects under threshold and marginal wall crossing.

\subsection{The c-map: From particles in 4d to instantons in 3d/4d} 
\label{from4dto3d}

There may seem to be an immediate problem with the idea that the topological string computes non-perturbative D-brane instanton effects, because it depends on the `wrong' moduli. Namely, one might have guessed that e.g. the A-model could describe D-brane instantons effects from topological A-branes, but the latter actually depend on complex structure moduli, which are visible only in the B-model! Equivalently, the A-model is usually related to the vector multiplets of the physical type IIA string theory, while non-perturbative D-brane instantons correct the hypermultiplet moduli space.

The solution to this puzzle is the so-called \emph{c-map}.
The c-map basically consists in dimensionally reducing a 4d $\mathcal{N}=2$ theory on a circle to 3d, T-dualizing, and decompactifying the T-dual to another 4d $\mathcal{N}=2$ theory. The c-map swaps type IIA and IIB theories, and the roles of vector and hypermultiplets. This provides a (to our knowledge, new) physical interpretation of the proposed S-duality of topological strings \cite{Nekrasov:2004js}, by which the A-model computes non-perturbative effects of the B-model, and viceversa.

The c-map involves a T-duality relating D-brane particles and instantons, and the corrections they induce, as already exploited in \cite{Ooguri:1996me,Seiberg:1996ns}. Starting in IIA (resp. IIB) string theory compactified on a CY threefold $X$, one can compute the quantum corrections to the vector-multiplet moduli space $\mathcal{V}_A$ (resp. $\mathcal{V}_B$) by computing one-loop diagrams of BPS particles arising from D-branes on holomorphic (resp. special lagrangian) cycles. Upon $\IS^1$ compactification to 3d, diagrams involving D-brane particles running along the $\IS^1$ can be regarded as instantons correcting the 3d vector multiplets. These can actually be dualized to 3d hypermultiplets parametrizing a quaternionic Kahler space $\mathcal{\hat V}_A$ (resp. $\mathcal{\hat V}_B$), fibered over the 4d vector moduli space with fibers given by the electric and magnetic Wilson lines along $\IS^1$. Upon T-duality and decompatification, $\mathcal{\hat V}_A$ (resp. $\mathcal{\hat V}_B$) become the 4d hypermultiplet moduli space of the T-dual IIB theory (resp. IIA). The D-brane particles map to 4d instantons from D-branes on holomorphic (resp. special lagrangian) cycles.

In next section we apply the c-map to relate loops of D2/D0-brane particles, arising naturally in the GV interpretation of the topological A-model, to D1/D(-1)-brane instanton corrections to hypermultiplet moduli space of type IIB compactifications. 

\subsection{The GV interpretation of the topological string}
\label{gvargu}

In this section we  show that the GV interpretation of the topological string provides a computational tool for D-brane instanton corrections to hypermultiplet moduli spaces in type IIB compactifications. More explicitly, in the GV interpretation \cite{Gopakumar:1998vy,Gopakumar:1998ii,Gopakumar:1998jq}
the A-model on $X$ relates to a one-loop diagram in M-theory on $X\times\IS^1$, involving BPS particles from M2-branes wrapped on holomorphic cycles on $X$, running with momentum along $\IS^1$. In the type IIA picture, one recovers a Schwinger diagram involving D2/D0-branes coupled to a graviphoton field strength background, and including non-perturbative effects from D2/D0-particle pair production. Using the c-map (further circle compactification to 3d and T-duality) these determine non-perturbative effects from D1/D(-1)-brane instantons in the T-dual type IIB on $X$.

The GV amplitude in M-theory on  $X \times S^1$ is determined by the GV invariants
$n_{\bf k}^r$, which provide the multiplicity of M2-branes wrapped on a holomorphic cycle in the class $[\gamma]=k_i[\gamma_i]$, with $SU(2)_L$ spin content $[(\frac 12)+2(0)]\otimes [(\frac 12)+2(0)]^r$. Here $[\gamma_i]$ are a basis of $H_2(X,\IZ)$. Then, the topological string partition function can be written as
\begin{equation}
F_{\rm top.}\, =\, \sum_{r,{\bf k}, p>0}\,  \frac{n_{\bf k}^r}{p}\,  \left(2\sin \frac{p\lambda}2\right)^{2r-2}\, \exp \, [\, -2\pi \, p\, k_i t_i]\,
\end{equation}
This partition function is derived by computing the one-loop Schwinger effect mentioned before directly in M-theory, in the presence of a constant, self-dual graviphoton background $F$. The perturbative part of this diagram reproduces the perturbative corrections to the vector multiplets of type IIA on $X$ computed by the topological string, which in the presence of the graviphoton background read
\beqa
S_{R^2,4d}\, =\, \int\, d^4x\, \sum_g \, F_g(t_i)\,  \lambda^{2g-2} R_{+}^{\, 2}\, =\,  \int\, d^4x\, F_{\rm top}(\lambda,t_i)\, R_+^{\, 2}
\eeqa
where the combination $g_s F=\lambda$ plays the role of topological string coupling. 
In addition, the GV formulation includes non-perturbative information from the D2/D0-brane particles. Through the c-map, the M-theory computation describes corrections from worldsheet and D1/D(-1)-brane instantons to the hypermultiplet moduli space of type IIB on $X$.

In the above discussion, different values of $r$ determine different kinds of corrections.
It is interesting to keep track of the label $r$ in this process and translate it to D-brane instanton language, both from the spacetime and brane worldvolume viewpoints. From the spacetime viewpoint, in M-theory on $X$, the label $r$ defines the spin content of the 5d M2-brane particle multiplet. In the spacetime Schwinger computation, a particle with spin $r$ introduces at least $r$ powers of the graviphoton field strength $F$, i.e. leads to a $(2r+2)$-derivative correction to vector multiplets. Via the c-map, it leads to a $(2r+2)$ derivative correction to hypermultiplets. From the brane worldvolume viewpoint, in M-theory on $X$, the label $r$ determines the number of fermion zero modes in the superparticle Quantum Mechanics (whose quantization leads to the spacetime spin content). From this viewpoint, the $F^{2r}$ correction in the Schwinger computation, arises from the coupling of the graviphoton field strength to the worldline fermion zero modes, and saturation of the latter in the path integral. Mapping the BPS particles to D-brane instantons in the T-dual type IIB, $r$ determines the number of fermion zero modes of the D-brane instanton (besides the four universal goldstinos, which relate to the center of mass half-hypermultiplet). From familiar D-brane instanton physics, it leads to a $(2r+2)$-derivative correction to the hypermultiplets, in agreement with the spacetime picture above.

As mentioned before, not much is known about higher-derivative F-terms for hypermultiplets, so we primarily focus on corrections to the hypermultiplet metric. These arise from the $r=0$ sector, namely D-brane instantons with four fermion zero modes, or D-brane particles in hypermultiplets. Geometrically, these correspond to M2-branes on $\IS^2$ 2-cycles, with multiplicity counted by the genus 0 GV invariants $n_{\bf k}^{(0)}$. In the IIA picture they become D2-branes on $\IS^2$'s with induced D0-brane charge, and in the IIB picture D1-brane instantons on $\IS^2$'s with induced D(-1)-brane charge.

\subsection{Threshold wall crossing}
\label{threshold-walls}

In this section we start exploring the implications of the above topological string connection for wall crossing of D-brane instantons. The above procedure produces non-perturbative D-brane instanton corrections, manifestly continuous throughout moduli space. This follows from the fact that GV BPS index is constant and has no wall crossing. This property, was already emphasized in \cite{Gopakumar:1998jq}, and is familiar and extensively used in the literature on BPS states. Thus this familiar statement provides an underlying explanation for the detailed conspiracies of D-brane instantons in multi-instanton processes, as we now discuss.

In the above discussion, D1/D(-1)-brane instantons play a prominent role. Since all instantons in this sector have mutually local charges, it seems natural to think  that the discussion is restricted to threshold stability walls. In this Section we indeed focus on this situation. However, the above system allows for a non-trivial discussion of marginal stability walls, as we discuss in an explicit example in Section \ref{marginal-walls}. 

There are indeed many examples of threshold walls in the D1/D(-1)-brane 
instanton sector. For instance, in the $X_3$ example in Section  \ref{nonpertacrosswalls}, with quiver in Figure \ref{quivers}a, a BPS D1/D(-1)-brane wrapped on an $\IS^2$ splits into two BPS sub-objects, described by two D1-branes on $\IS^2$ touching at a point. Despite the splitting, the relevant corrections to the hypermultiplet moduli space metric are continuous. From the microscopic instanton viewpoint, this follows from a tricky 2-instanton process; from the topological string viewpoint, it is manifest from the continuity of the GV invariants. It is interesting to relate this two viewpoints more explicitly. In the topological string computation, the particle arising from the wrapped M2-brane (or D2/D0-branes) actually splits into two classical particles. However, the 2-particle system has a quantum BPS bound state at threshold, as one can easily show from the corresponding quiver quantum mechanics, i.e. the reduction to $0+1$ dimensions of the 4d $\NN=1$ theory described by the quiver in Figure \ref{quivers}a. Following \cite{Denef:2002ru}, the problem amounts to the computation  of the Euler characteristic of the moduli space of the quiver gauge theory. The moduli space is defined by the vanishing of the worldvolume D- and F-term potential, namely
\beqa
|\Phi_{12}|^2\, -\, |\Phi_{21}|^2\, =\, 0 \nonumber \\
\Phi_{12}\Phi_{21}\Phi_{12}\, =\, 0 \nonumber \\
\Phi_{21}\Phi_{12}\Phi_{21}\, =\, 0 
\label{modulispace}
\eeqa
The moduli space is a point $\Phi_{12}=\Phi_{21}=0$, or rather the $\IR^3$ of the relative positions in 4d. The Euler characteristic (with compact support) is $\chi({\cal{M}})=1$, leading to a single BPS bound state \footnote{The existence of these threshold states has been shown from alternative standpoints. For instance by using geometric quantization of the two-particle phase space \cite{deBoer:2008fk,deBoer:2008zn} in multi-center supergravity solutions given by split attractor flows \cite{Denef:2000nb,Denef:2007vg,Collinucci:2008ht}, in a quite generic example (binding of a D0-brane with a D6/${\ov{\rm D6}}$-brane system).}. At the threshold wall, the 2-particle bound state running in the Schwinger loop reproduces the 2-instanton process of the T-dual side. 

\subsection{Marginal wall crossing}
\label{marginal-walls}

Even though our discussion is restricted to D1/D(-1)-brane charges, it still provides non-trivial information about the behaviour of non-perturbative effects across certain marginal stability walls. Indeed, in certain models there are regions in moduli space where D1/D(-1) instantons are marginally unstable against decay into pairs of D3-\aD3 brane-antibrane instantons (with net induced D1/D(-1)-brane charge), as we later show in an explicit example. The continuity of the corresponding non-perturbative effects is however manifest in terms of the topological string computation, since the latter is determined by the GV invariants, which are continuous throughout moduli space. From the M-theory perspective, the continuity of the non-perturbative effect would follow from the fact that the D1 $\to$ D3-\aD3  wall crossing does not lift to M-theory. Indeed, it would imply a decay of 5d M2-brane particles into M5 brane-antibrane pairs, which actually cannot produce 5d particles. Hence the M2-brane particles are stable throughout moduli space
 \footnote{We thank E. Witten for this interpretation.}.
Note that there is no claim that the topological string is capturing the non-perturbative effects of individual D3-or anti-D3-brane instantons, but rather that it captures the information of the (non-BPS) 2-instanton 
process (which does not carry a net D3-brane charge) relevant to the continuity of the non-perturbative terms. Inclusion of general D-brane charges is discussed in Section \ref{general-charges}. 

\begin{figure}
\begin{center}
  \includegraphics[width=11cm]{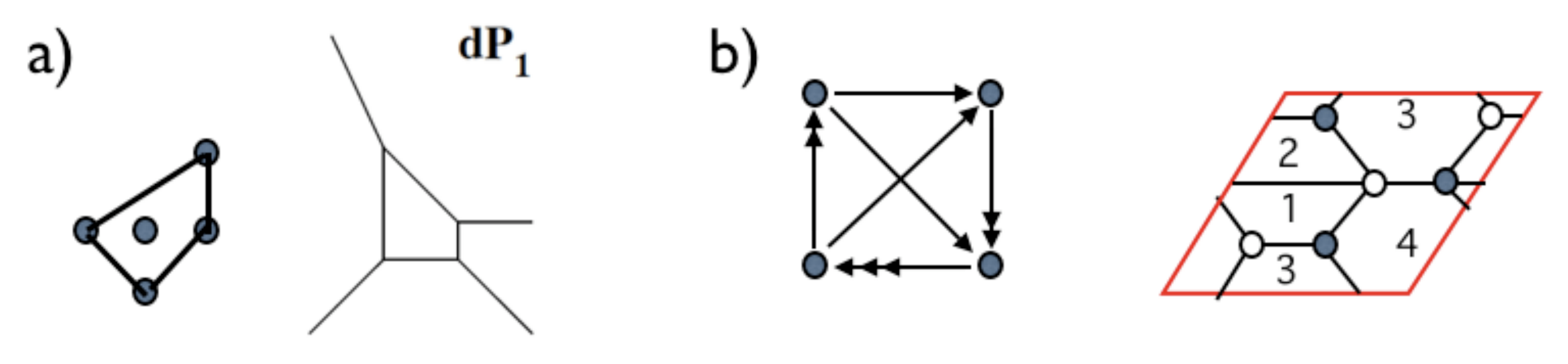}
  \caption{\small a) Toric diagram and web diagram for the complex cone over $dP_1$. For clarity, in the web diagram we show the collapsed cycles with some finite but small size. b) The quiver and dimer diagrams for the theory..}
  \label{diagrams-dp1}
\end{center}
\end{figure}

For concreteness let us present an example of marginal decay of a D1 into a D3-\aD3 pair. In order for the latter to be BPS, the system must be realized in the small volume regime, where BPS phases have $\alpha'$ corrections. Our example is based on D-branes at toric singularities, see \cite{Kennaway:2007tq} for a review. Consider the complex cone over $dP_1$, whose toric and web diagrams are shown in Figure \ref{diagrams-dp1}a. The set of BPS D-branes in this geometry correspond to `fractional' D(-1)-brane instantons, whose properties can be encoded in a quiver diagram and dimer diagram, shown in Figure \ref{diagrams-dp1}b. In the latter, faces in the tiling correspond to gauge groups, edges correspond to bi-fundamental chiral multiplets, and nodes correspond to worldvolume superpotential terms. This diagram is particularly useful in discussing partial resolutions of singularities.

By blowing up and going to large volume, these fractional branes can be described as geometric D3 or \aD3-branes wrapped on the exceptional $dP_1$ and carrying worldvolume gauge bundles inducing D1/D(-1) -brane charges. The charges of such an object $F$ are encoded in the  Chern character 
\beqa
{\rm ch}(F)\, =\, ({\rm rk} (F), c_1(F),c_2(F))
\eeqa
The large volume charges can be determined (up to monodromy in the path to large volume) from exceptional collections, see \cite{Hanany:2006nm} for a constructive recipe for toric singularities. For the $dP_1$ theory, a convenient choice is
\beqa
& {\rm ch}(F_1)  =  (1,H-3E,0) \quad \quad \quad\quad
& {\rm ch}(F_2)  =  (1,-E,-\frac 12) \nonumber \\
&{\rm ch}(F_3)  =  (-1,-H+2E,-\frac 12) \quad
&{\rm ch}(F_4)  =  (-1,2E,0)
\eeqa
where $H$, $E$ are the hyperplane and exceptional classes in $dP_1$. It is easy to see that the DSZ product for the above large volume D-branes reproduces the quiver of the $dP_1$ theory shown above. In the large volume limit these branes have different BPS phases, but they are aligned in the singular geometry. 

\begin{figure}
\begin{center}
  \includegraphics[width=11cm]{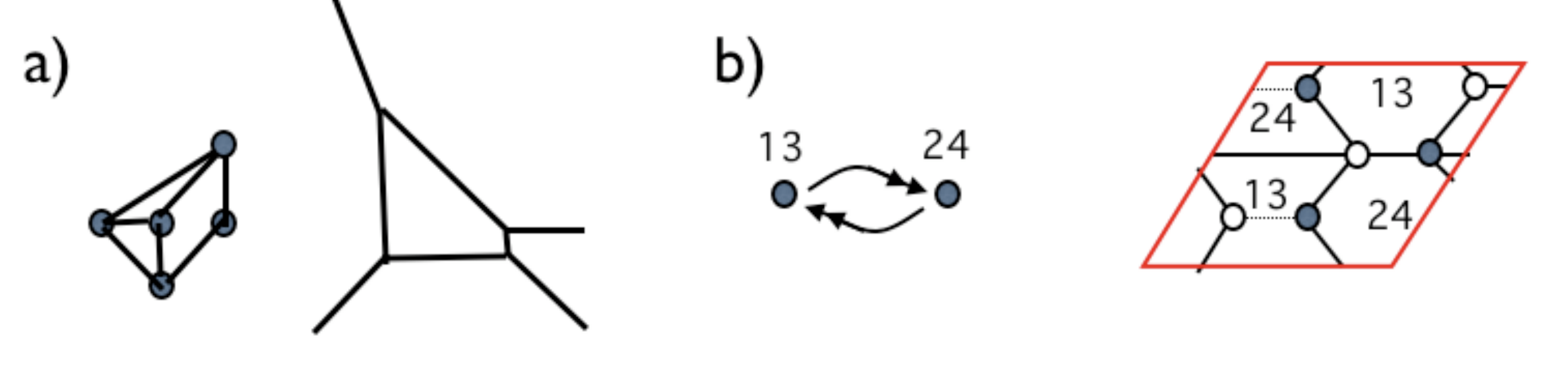}
  \caption{\small a) Toric diagram and web diagram for the partial blow-up to a conifold singularity. For clarity, in the web diagram we show the collapsed cycles with some finite but small size, while the blown-up cycles are shown with finite and large size. b) The quiver and dimer diagram for the resulting conifold singularity. They are obtained from the $dP_1$ ones by recombining the nodes 1 with 3, and 2 with 4.}
  \label{diagrams-coni}
\end{center}
\end{figure}

The complex cone over $dP_1$ admits a partial blow-up to a conifold singularity, with $E$ remaining as the collapsed 2-cycle in the conifold tip. This is shown in the toric and web diagrams in Figure \ref{diagrams-coni}a. The blow-up Kahler modes couple as Fayet-Illiopoulos terms to the world-volume of the D-branes, forcing them to form bound states, which are BPS in the conifold geometry. In our example the dimer shows, figure \ref{diagrams-coni}b, a recombination of branes 1 with 3, and 2 with 4. In large volume language, the charges of these recombined branes are
\beqa
{\rm ch}(F_{13}) =  (0,-E,-\frac 12) \quad\quad\quad\quad
{\rm ch}(F_{24})  =  (0,E,-\frac 12)
\eeqa
namely branes 1 and 3 combine into a D1-brane wrapped on the $\IS^2$ of the conifold, with induced fractional D(-1)-brane charge, and the branes 2 and 4 combine into a D1-brane wrapped with opposite orientation (and same induced D(-1)-brane charge). This is the familiar geometric description of fractional D-branes at a conifold. 

We are now ready to discuss the marginal stability wall. Consider the $dP_1$ singularity partially blown up to a conifold, and consider a D-brane instanton associated to the node 24, namely a D1-brane wrapped on the $\IS^2$ and with induced D(-1)-brane charge. Moving in moduli space to the blown-down $dP_1$ singular configuration, the D-brane splits into a pair of D-branes associated to the nodes 2 and 4 in the $dP_1$ quiver. These have mutually non-local charges, and describe a marginal stability decay, with quiver given by setting $n_1=n_3=0, n_2=n_4=1$ in the general $dP_1$ quiver, namely
a $U(1)\times U(1)$ theory with a single bifundamental chiral multiplet. Hence the singular $dP_1$ geometry is sitting at a marginal stability wall for the D1-brane on $E$ to decay to a D3-\aD3 pair. 

\section{From M-theory to 3d and type IIB instantons}
\label{3daction}

Higher derivative corrections to hypermultiplets are not well understood in general (see \cite{Michel:2007vh} for a partial analysis of higher derivative corrections), so most analysis focus on the computation of hypermultiplet moduli space metrics \cite{Rocek:2005ij,RoblesLlana:2006ez}. We will also focus in this case in the explicit computation in this Section.

The analysis of \cite{Ooguri:1996me} and \cite{Seiberg:1996ns} can be easily generalized, and applied to the computation of type IIA or type IIB D-brane instantons 
in mutually local sectors. In this Section we elaborate on the computation of type IIB D1/D(-1)-brane instanton effects by computing a one-loop diagram of D2/D0-brane particle states in type IIA compactified to 3d.  

Actually, this computation can be lifted to M-theory, displaying a structure identical in spirit to the GV interpretation of the topological string. Namely, we consider M-theory compactified on $X$, and subsequently compactified to 3d on $\IT^2$. The amplitude of interest is a one loop diagram of 5d BPS states, with momentum along the $\IT^2$. Such BPS particles are supersymmetric gravitons, and wrapped M2-branes, with multiplicities counted by the GV invariants.  As mentioned, states with different spin label $r$ contribute to different terms in the 3d theory, hence corrections to the hypermultiplet moduli space from wrapped M2-branes are related to the genus zero GV invariants.

The type IIB non-perturbative brane instanton effects are recovered by taking the limit of zero area of the $\IT^2$, keeping its complex structure $\tau$ constant (which becomes the complex string coupling in the IIB side). Notice that the M-theory picture has a manifest $SL(2,\IZ)$ invariance with respect to the $\IT^2$. This will translate to the $SL(2,\IZ)$ invariance of the non-perturbative effects on the IIB side. Namely, the M-theory computation reproduces a sum over $(p,q)$ string instanton effects, or equivalently over general D1/D(-1)-brane instantons, including D1-branes with non-trivial worldvolume fluxes. Rather than taking the intermediate step of the type IIA limit (which would correspond to a particular limit of weak coupling $\tau \to i\infty$) we work at general $\tau$ and recover the full result. 

General expressions for these non-perturbative effects in 4d IIB models on $X$ have been obtained in \cite{RoblesLlana:2006is,RoblesLlana:2007ae}, by starting with worldsheet instantons and imposing $SL(2,\IZ)$ invariance. The M-theory perspective provides a simple derivation of the result in a manifestly $SL(2,\IZ)$ invariant formulation; it also explains the observation that the result depends on $X$ only through its Euler characteristic and the genus 0 GV invariants. 
The main purpose of carrying out the computations that follow will not be to obtain new formulae \emph{per se}, but to establish the natural connection between non-perturbative effects and the topological string.
\subsubsection*{The graviton piece}

Let us describe the contribution arising from 5d gravitons. These are 11d gravitons, truncated to their zero modes in $X$, and running with arbitrary momenta in the $\IT^2$ (and the non-compact 3d). The computation can be regarded as a truncation (to the zero mode sector in $X$) of the computation in \cite{Green:1997as} of D(-1)-brane instanton contributions to a certain $R^4$ term for IIB on $\IS^1$ from graviton one-loop diagram in M-theory on $\IT^2$. Indeed, the reduction of $R^4$ terms on a CY $X$ provide corrections to the hypermultiplet moduli space metric, see e.g. \cite{Antoniadis:1997eg}.
Morally, the $R^4$ term with three curvatures insertions along $X$ (producing a numerical factor given by the Euler characteristic) leads to a correction to the 3d Einstein term, which can be recast as a correction to the hypermultiplet metric. 

The 5d graviton BPS states are obtained by considering the quantum ground states of gravitons in $X$. Accounting for boson-fermion cancellations, their index is given by  the Euler characteristic of $X$, $\chi_E(X)=2\,(h^{1,1}-h^{2,1})$. The correction to the hypermultiplet moduli space metric is obtained from a 1-loop diagram of these states, with one insertion of curvature tensor, which we will perform in the Schwinger \emph{proper time formalism}. The latter just contributes one power of the proper time parameter, i.e. the one-loop calculation is of the form
\begin{equation}
A \propto \Tr \left( \int \frac{ds}{s} \,s\, e^{-s\,H} \right) = \int d({\rm momenta})\int \frac{ds}{s} \,s\, e^{-s\,H}\,,
\end{equation}
where $s$ is the proper time parameter of the graviton, and the positive power of $s$ corresponds to one insertion of the curvature tensor. The momenta include the 3d non-compact directions and the $\IT^2$ directions.

More precisely, the one-loop graviton amplitude with one graviton insertion reads
\beqa
A_{g}\, =\, -\frac{\chi_E(X)}{\pi\,\mathcal{V}_2}\, \int \, d^3p\, \int_0^{\infty} ds\,\,\sum_{\{l_I\}} e^{-s\,\left({\bf p}^2+G^{IJ}\,l_I\,l_J\right)}\, ,
\eeqa
where $G^{IJ}$ denotes the inverse metric on the $\IT^2$. Integrating over 3d momenta $p$, and performing a Poisson resummation on the discrete momenta $l_I$, we have
\begin{align}
A_{g} = - \chi_E(X)\,\pi^{5/2}\, 
\int_{0}^{\infty}d\tilde s \,{\tilde s}^{1/2}\, e^{-\tilde s\,\pi^2\,G_{IJ}\,k^I\,k^J}\,,
\end{align}
where $\tilde s = 1/s$, and $\mathcal{V}_2$ is the $\IT^2$ area. For a $\IT^2$, $G_{IJ}\,k^I\,k^J = |k_1+\tau\,k_2|^2\,\mathcal{V}_2/\tau_2$ with $\tau=\tau_1+i\tau_2$ the $\IT^2$ complex structure. Leaving out the $\vec{k}=\vec{0}$ term and performing the integral over $\tilde s$, we obtain
\beqa
A_{g} =  -\frac{\chi_E(X)}{2\,\mathcal{V}_2^{3/2}}\,
\sum_{(k_1, k_2) \neq (0,0)} \frac{\tau_2^{3/2}}{|\, k_1+\tau\,k_2\, |^3}\, \equiv \, -\frac{\chi_E(X)}{2\,\mathcal{V}_2^{3/2}}\,E(\tau,3/2).
\eeqa

\subsubsection*{The M2-brane piece}

We now compute the one-loop diagram with one external graviton leg and with the particle in the loop given by an M2-brane wrapped on a genus zero holomorphic 2-cycle $C \in H_2(X,\mathbb{Z})$ of $X$. The effective particle has a five-dimensional mass $m_C \propto Vol(C) = \int_C J$, where $J$ is the K\"ahler form, and its worldline action has a coupling to a gauge field $A_\mu = \int_{C} C^{(3)}$
\begin{equation}
\Delta S = \int_{\rm w.l.} A_\mu\,\dot{X}^\mu\,.
\end{equation}
It shifts the momentum of the particle, so that the effective worldline Hamiltonian is 
\begin{equation}
H = ({\bf p+A})^2+G^{IJ}(l_I+A_I)\,(l_J+A_J)+m_C^2\,.
\end{equation}
The one-loop amplitude (including the factor of $s$ from the external graviton leg) is
\begin{align}
A_{\rm M2}&= \frac{1}{\pi\,\mathcal{V}_2}\,\int d^{3}\,p\,\int_0^{\infty} ds\,\sum_{\{l_I\}} e^{-s\,\left({\bf (p+A)}^2+G^{IJ}(l_I+A_I)\,(l_J+A_J)+m_C^2\,,\right)}\,.
\end{align} 
By redefining the continuous momenta, the shift by ${\bf A}$ has no effect in their integral, which is straightforward. For the discrete momenta, the Wilson line has an effect in the subsequent Poisson resummation, as follows:
\begin{align}
A_{\rm M2}&= 
\pi^{5/2}\,\int_{0}^{\infty} ds\,s^{-5/2}\,\sum_{(k_1, k_2) \neq (0,0)}e^{2\,\pi\,i\,k^I\,A_I}\,e^{-\pi^2\,G_{IJ}\,k^I\,k^J/s-s\,m_C^2}\\
&=\pi^{5/2}\,\,\int_{0}^{\infty} d\tilde s\,\tilde s^{1/2}\,\sum_{(k_1, k_2) \neq (0,0))}e^{2\,\pi\,i\,k^I\,A_I}\,e^{-\tilde s\, \pi^2\,G_{IJ}\,k^I\,k^J-m_C^2/\tilde s}\,.
\end{align}
The integral over $\tilde s$ can be evaluated by redefining the variable as $\tilde s \rightarrow X^2$. The integral then has the form:
\begin{align}
&= e^{-2\,\sqrt{A\,B}}\,\frac{\sqrt{\pi}}{4\,A^{3/2}}\, (1+2\,\sqrt{A\,B})\, ,
\end{align}
with $A = \pi^2 k \cdot k = \pi^2\,G_{IJ}\,k^I\,k^J\, =\, \pi^2\, |k_1+\tau k_2|^2\,\mathcal{V}_2/\tau_2$ and $B=m^2_C$. To write the final result, we define the effective 3d `mass' (Euclidean action) of the resulting instanton as $m^{(3)}_C \equiv \sqrt{\mathcal{V}_2/\tau_2}\,m_C$.
\begin{align}
A_{\rm M2} &= \frac{1}{\mathcal{V}_2^{3/2}}\,\sum_{(k_1,k_2)\neq (0,0)} 
\,\frac{\tau_2^{3/2}}{|k_1+\tau k_2|^{3}}\, (1+2\,\pi\,|k_1+\tau k_2|\,m^{(3)}_C)\, 
e^{-2\,\pi\,(|k_1+\tau k_2| \, m^{(3)}_C\,  - \, i\,k_1\,A_1\, -\, i\,k_2\,A_2\,)}\,.
\end{align}
\vskip 2mm
The calculation performed above was for a membrane wrapped on a fixed, rational (genus zero) holomorphic curve in a class $[C] \in H_2(X, \mathbb{Z})$. The full amplitude is a sum over all possible curve classes. Let $[\gamma_a]$ denote a basis of $H_2(X, \mathbb{Z})$, and expand any curve $[C]=k_a[\gamma_a]$. The number of BPS M2-branes wrapped on a genus zero representative of the class $[C]$ is given by the genus zero GV invariant $n_{\bf k}^{(0)}$. Dropping overall volume factors, the total correction to the hypermultiplet moduli space metric from such M2-branes is
\begin{equation} \label{membranepiece}
A_{\rm M2 \ total} = \frac{1}{4\,\pi} \sum_{{\bf k}} n_{k_a}^{(0)}\, 
\sum_{(m, n)\neq (0,0)} \,\frac{\tau_2^{3/2}}{\,|m+\tau n|^{3/2}}\, (1+2\,\pi\,|m+\tau n|\,k_a\,t^a)\,  e^{-S_{m, n}}\quad 
\end{equation}
where we have introduced
\beqa
S_{m, n}\, =\, 2\,\pi\,(\, |m+\tau n| \, k_a\,t^a\,  - \, i\,m \,b^a\, -\, i\,n\, c^a\,)
\label{M2action}
\eeqa
the action of an euclidean M2-brane particle wrapped on the $(m, n)$ 1-cycle on $\IT^2$, and we have normalized the K\"ahler parameters $t^a$ such that
\begin{equation}
\int_{[\gamma_a]}J = t^a\,\sqrt{\frac{\mathcal{V}_2}{\tau_2}}\,.
\end{equation}
For later convenience, we have named our Wilson lines as follows:
\begin{equation}
\int_{[\gamma_a]\times S^1_{i=1,2}} C^{(3)} = (b^a, c^a)\,,
\end{equation}
where the two components represent Wilson lines along the first and the second circle of the $\IT^2$, respectively.

By shrinking one of the torus 1-cycles one can recover an interpretation in type IIA on $X\times \IS^1$. From this perspective, the above amplitude contains in particular the contributions from worldsheet instantons (M2-branes wrapping the M-theory circle) and from D2/D0-brane bound states (from M2-branes wrapped on the M-theory circle, and carrying momentum along it). 
In fact, by taking such a IIA limit, we can bring the formula \eqref{membranepiece} to a form that is reminiscent of the Gopakumar-Vafa formula in \cite{Gopakumar:1998vy}. By taking, for instance, $\tau_2 \rightarrow \infty$, and dropping the sum over $n$, we see that the second term in the expansion becomes
\begin{align} \label{dilogmembrane}
 & \frac{1}{2} \sum_{{\bf k}} n_{k_a}^{(0)}\, 
\sum_{(n)\neq 0} \,\frac{1}{\,n^{2}}\,e^{-2\,\pi\,(\, k_a\,t^a\,  -\, i\, c^a\,)\,|n|}\\  
&=   \sum_{{\bf k}} n_{k_a}^{(0)}\, 
{\rm Li_2}(e^{-2\,\pi\,(\, k_a\,t^a\,  -\, i\, c^a\,)\,n})\,,
\end{align}
where we have introduced the \emph{dilogarithm} ${\rm Li}_2(x)=\sum_{n=1}^\infty\, x^k/n^2$.
The appearance of such dilogarithms is reminiscent of the structure in \cite{ks}. In section
\ref{general-charges} we will consider the role of dilogarithms in this connection.

\subsubsection*{The type IIB instanton interpretation}
\label{iibinstantons}

The above results can be related to type IIB 4d brane instantons by shrinking the $\IT^2$ keeping the complex structure $\tau$ fixed. This process maps the manifest geometric $SL(2,\IZ)$ invariance of M-theory to the S-duality group of IIB. In order to do so, it is convenient to notice that the above structures clearly reproduce the type IIB result in \cite{RoblesLlana:2006is,RoblesLlana:2007ae}, 
which we review here for completeness. In the off-shell $\NN=2$ formalism of these references, the hypermultiplet metric is encoded in a single function, the tensor potential, 
which contains a classical piece $\chi_{\rm cl.}$, plus contributions related to D(-1)-brane and $(p,q)$ 1-brane instantons
\beqa
\chi\, =\, \chi_{\rm cl.}\, +\, \chi_{\rm D(-1)}\, +\, \chi_{\rm 1-brane}
\eeqa
with
\beqa
\chi_{\rm cl} & = &4\, r^0 \, \tau_2^2 \, \frac{1}{3!} \,
\kappa_{abc} \, t^a \, t^b \, t^c\ , \nonumber \\
\chi_{\rm D(-1)} & = &\frac{r^0 \tau_2^{1/2}}{2 (2 \pi)^3}\,
\chi_E(X) \, \sum_{(m,n)\neq (0,0)}\, \frac{\tau_2^{3/2}}{|m\tau +
n|^3}\ , \\
\chi_{\rm 1-brane} & = &- \frac{r^0 \tau_2^{1/2}}{(2 \pi)^3}\,
\sum_{\bf k} n_{k_a}^{(0)}  \sum_{(m,n)\neq (0,0)} \frac{\tau_2^{3/2}}
{|m\tau + n|^3}\, \big( 1 + 2 \pi |m\tau + n|\, k_a t^a \big) \,
e^{-S_{m,n}}\ .\nonumber
\label{vandoreniib}
\eeqa

Here $\kappa_{abc}$ are the classical triple intersection form on the CY, and $r^0$ is a scalar in the tensor multiplet to render the tensor potential of appropriate degree, see \cite{RoblesLlana:2006is,RoblesLlana:2007ae} for notation and further details. The $n_{\bf k}^{(0)}$ denote the genus zero GV invariants. Also,
\beqa
  S_{m,n} = 2\pi k_a \,(\, |m\tau + n|\, t^a - \ i m\, c^a - \ i n\, b^a\, )\ ,
\label{stringaction}
\eeqa
is the tension of a type IIB $(p,q)$ string wrapped $s={\rm gcd}(m,n)$ times around a holomorphic cycle, with $(p,q)=(m,n)/r$, and agrees with the action (\ref{M2action}) of an M2-brane wrapped on the 2-cycle on $X$ times a 1-cycle on $\IT^2$. The axion scalars $b^a$, $c^a$ are integrals of the IIB NSNS and RR 2-forms over the 2-cycle $\gamma_a$ correspond to the 3-form integrals in the M-theory picture.

As mentioned, the M-theory derivation of these known results explains naturally their structure, in particular the $SL(2,\IZ)$ invariance, and the apperance of the topological invariants $\chi_E(X)$ and $n_{\bf k}^{(0)}$. But their main interest is the fact that this derivation proves a strong connection between non-perturbative effects and topological strings. It would be interesting to generalize this analysis to higher derivative terms and shed some light on other protected couplings of 4d $\NN=2$  hypermultiplets.

\section{Non-perturbative effects and wall crossing from matrix model instantons}
\label{npdefinitions}

In our previous discussion, the computation of the non-perturbative terms in the topological string partition function requires an a priori knowledge of the set of BPS wrapped brane states. However, there are frameworks which, for certain CY models, have been proposed to provide a non-perturbative definition of the topological string, in the sense of \cite{Marino:2008ya}. An example, on which we focus in this Section, is given by matrix models \footnote{Another kind of non-perturbative definition is provided by Chern-Simons theories (which incidentally also admit a description in terms of matrix models \cite{Marino:2002fk}), see \cite{Marino:2004eq,Marino:2005sj} for reviews. Non-perturbative effects in Chern-Simons models have been explicitly related to Schwinger processes of the physical theory \cite{Gopakumar:1998vy}.  
 }. In these frameworks, both the CY geometry and all its relevant data emerge from the defining system. We center on topological string B-models defined by  Dijkgraaf-Vafa matrix models \cite{Dijkgraaf:2002fc,Dijkgraaf:2002vw,Dijkgraaf:2002dh} \footnote{The B-model on many local CY (for instance the mirror of toric CY's) has been solved \cite{Bouchard:2007ys} by using the invariants of \cite{Eynard:2007kz} on certain spectral curves. The description of these models in terms of matrix models is however not explicitly known, so the knowledge of non-perturbative effects is more limited. We thus focus on topological string models defined in terms of {\em bona fide} matrix models.}, for which non-perturbative effects from instantons have been recently discussed in \cite{Marino:2007te,Marino:2008vx,Eynard:2008he}. 

In this Section we propose a new physical interpretation for these processes, in terms of the mirror B-model side of our previous discussions. Namely we propose that matrix model instantons correspond to non-perturbative effects from D3-brane particles in the physical type IIB theory, which can subsequently be related through the c-map to D2-brane instanton effects in the T-dual IIA theory. This can be shown very explicitly for the conifold \cite{Pasquetti:2009jg}. We also study a concrete set-up in which these physical D-branes cross walls of stability, but their non-perturbative effect (given by the matrix model instanton amplitude) is continuous. 

\subsection{Review of matrix models}

Let us start by briefly reviewing the Dijkgraaf-Vafa matrix model \cite{Dijkgraaf:2002fc,Dijkgraaf:2002vw,Dijkgraaf:2002dh}.
The Dijkgraaf-Vafa correspondence relates the topological B-model on 
a certain CY (to be discussed later) with the matrix model defined by the partition function
\begin{equation}
Z=\frac{1}{{\rm Vol}(U(N))}\int dM \exp\left(-\frac{1}{g_s}{\rm Tr}\, W(M)\right).
\end{equation}
The variables $M$ of the model are $N\times N$ matrices, and we will consider a polynomial 
potential $W(M)$ of degree $(n+1)$. Clearly, the action is invariant under 
the gauge symmetry $M\to UMU^\dag$, which can be used to diagonalize the matrices and recast the partition function as an integral over the $N$ eigenvalues $\lambda_i$:
\begin{equation}\label{partition1}
Z=\frac{1}{N!}\int \prod_i \frac{d\lambda_i}{2\pi} e^{N^2S_{eff}(\lambda)},
\end{equation}
where the effective action is given by
\begin{equation}
S_{eff}(\lambda)=-\frac{1}{tN}W_{eff}(\lambda)= -\frac{1}{tN}\sum_{i=1}^N W(\lambda_i)+\frac{2}{N^2}\sum_{i<j}\log{|\lambda_i-\lambda_j|}.
\end{equation}
Here $t=g_s N$ is the usual 't Hooft parameter, and the logarithmic term arises
from the Jacobian upon diagonalization. It acts as a repulsive force between the eigenvalues, which behave as particles in the potential $W_{eff}(\lambda)$ \footnote{The above description corresponds to the hermitean matrix model, where eigenvalues are forced to lie in the real line. We will continue our discussion in the holomorphic matrix model \cite{Lazaroiu:2003vh}, where eigenvalues live on general contours in the complex planes.}.

The saddle points of the model would correspond (once we relate it to the topological string) to distributions of the eigenvalues among the $n$ extrema of the potential $W$. Their repulsion actually makes the eigenvalues distribute around the extrema, filling $n$ domains or cuts ${\mathcal C}_i$ around them. The contribution to the partition function from a saddle configuration with $N_i$ eigenvalues in ${\cal {C}}_i$  is
\begin{equation}\label{partition2}
Z(N_1,\ldots, N_n)=\frac{1}{N_1!\ldots N_n!}\int_{\lambda^1_{k_1}\in{\cal{C}}_1}\ldots \int_{\lambda^n_{k_n}\in{\cal{C}}_n}\,  \prod_i \frac{d\lambda_i}{2\pi} e^{N^2S_{eff}(\lambda_i)},
\end{equation}
We are actually interested in the planar limit of the matrix model, where $N\to \infty$, keeping the 't Hooft coupling and the filling fractions $\nu_i=N_i/N$ finite. Around a saddle configuration, the matrix model can be solved in terms of the so-called resolvent function $y(x)$ which has a useful expression in terms of the endpoints of the cuts $x_i$:
\begin{equation}
y(x)=W'(x)-\oint_{\mathcal{C}}\frac{dz}{(2\pi i)}\frac{W'(z)}{x-z}\prod_{i=1}^{2n}\left(\frac{x-x_i}{z-x_i}\right)^{\frac{1}{2}},
\end{equation}
where ${\mathcal C}= \cup_{i=1}^n{\mathcal C}_i$ (see \cite{Marino:2004eq} for further details). 
We can perform the integral above by deforming the contour of integration to infinity and redefining $z\to 1/z$. Using also the polynomial form of the potential $W(x)$ the resolvent can be compactly written as 
\begin{equation}\label{spectral}
y(x)=\prod_{i=1}^{2n}(x-x_i)^\frac{1}{2}.
\end{equation}

The Dijkgraaf-Vafa correspondence relates this matrix model to the topological B-model on a non-compact CY $X$ defined by the equation
\begin{equation}\label{CY}
uv+y^2+W'\,^2(x)+f(x)=0,
\end{equation}
where the function $f(x)$ is polinomial of degree $n-1$ whose coefficients, whose details we will not need, depend on the filling fractions $\nu_i$. The basic geometry of $X$ is encoded in the Riemann surface at $uv=0$, namely (\ref{spectral}). This is a double cover of the complex $x$-plane with cuts ${\cal{C}}_i$. In particular, the 3-cycles of $X$ can be obtained by fibering $\IS^2$'s (parametrized by $u,v$) over non-trivial 1-cycles in the spectral curve. Namely, there are $a$-cycles, which can be obtained from 1-cycles $a_i$ surrounding the cuts, and their dual $b$-cycles which can be obtained by fibering over the non-compact 1-cycles $b_i$ stretching from the $i^{th}$ cut to infinity.
The holomorphic 3-form on $X$ is $\Omega= du\wedge dv \wedge dz/y$, and its restriction to the 3-cycle provides the $\NN=2$ BPS phase for a wrapped D-brane. By integrating over the fibers, the BPS phase can be obtained from the restriction to the 1-cycle in the Riemann surface of the 1-form
\beqa
\omega\, =\, y\,dx
\eeqa

\subsection{A new physical interpretation for matrix model instantons}

In the matrix model there are instanton effects around each saddle point. They correspond to tunneling of an eigenvalue from the cut $i$ to the cut $j$.  These instantons are roughly speaking weighted by $e^{-N}$, hence are non-perturbative in the coupling $g_s$.

The action for such instanton can be obtained by directly comparing the partition functions for both saddle points \cite{Marino:2008vx}
\beqa
\frac{Z(N_1,\ldots, N_i,\ldots, N_j,\ldots, N_n)}{Z(N_1,\ldots, N_i-1,\ldots, N_j+1,\ldots, N_n)} \simeq e^{-S_{\rm inst.}} \quad ; \quad S_{\rm inst.} \, =\, \frac{1}{g_s} \int_{C_{ij}} \, y(x)\, dx\quad 
\eeqa
where $C_{ij}$ is a path from the $i^{th}$ to the $j^{th}$ cut. The instanton action is independent of the path chosen. Such instantons have been shown to control the large order behaviour of the matrix model perturbative series around the saddle point
\cite{Marino:2007te}.

Let us consider the interpretation of these matrix model instantons in the physical theory.
The action of a matrix model instanton is related to the volume of a 3-cycle of the CY, given by a linear combination of the $b$-type 3-cycles, with $\IS^3$ topology. This suggests that the non-perturbative contributions are related to effects of 4d BPS particles 
\footnote{In \cite{Dijkgraaf:2002fc} a different interpretation was used, in terms of domain walls from D5-branes wrapped on the 3-cycle, interpolating between two vacua with different distribution of background gauge D5-branes. This interpretation is however not consistent with the fact that the number and distribution of eigenvalues does not correspond to the number and distribution of background gauge D5-branes.} from D3-branes wrapped on the 3-cycle $C_{ij}$. Running the logic of the previous sections, these BPS D-brane particle effects can be interpreted as 4d D2-brane instanton effects \footnote{In \cite{GarciaEtxebarria:2008iw} certain D-brane instantons of the physical theory were shown to be captured by the matrix model. It would be interesting to relate this effect to our discussion.} in the T-dual type IIA compactification on $X$.

Notice the interesting fact that the matrix model instantons only provide D-brane instantons  wrapped on compact $b$-cycles (paths joining different cuts), but not on the $a$-cycles (paths surrounding the cuts). This nicely matches our discussion (in the GV interpretation of the A-model) that topological strings naturally encode the information from D-brane objects of mutually local charges.

\subsection{Wall crossing in matrix models}

The above physical interpretation suggests that the non-perturbative effects in the physical theory computed from the matrix model instanton are completely insensitive to the microscopics of the underlying D-brane instanton. In fact, as one moves in moduli space a given D-brane instanton (say a D-brane wrapping the 3-cycle $C_{13}$) of the physical theory may split into components (say D-branes wrapping $C_{12}$ and $C_{23}$). In the matrix model the non-perturbative effect of either system is described as an eigenvalue jump between cuts ${\cal {C}}_1$ and ${\cal{C}}_3$. The field of forces in the matrix model is conservative, hence the matrix model instanton effect is independent of the path chosen for the eigenvalue.

Let us consider an explicit example in the simple case of a 3-cut matrix model, with cuts on the real axis, labeled 1,2,3 from left to right. There is an antiholomorphic involution $x\to ix$ which ensures that any path along the real axis has constant BPS phase. Hence in this situation there are matrix model instantons associated to D2-branes on $C_{12}$ and on $C_{23}$. The matrix model instanton taking an eigenvalue from cut ${\cal{C}}_1$ to cut ${\cal{C}}_3$ is actually related to a 2-instanton process in terms of D-branes wrapped on $C_{12}$ and $C_{23}$. Now consider changing one real parameter $\delta$, which controls the position of cut 2 away from the real axis. For large enough $\delta$ the model turns effectively into a 2-cut model, with only cuts ${\cal{C}}_1$, ${\cal{C}}_3$, still on the real axis. As we argue below, in the process the BPS 3-cycles $C_{12}$ and $C_{23}$ recombine into a BPS 3-cycle $C_{23}$, thus leading to a wall of BPS stability for the corresponding D-brane instantons in the physical theory. 

\begin{figure}
\begin{center}
  \includegraphics[width=10cm]{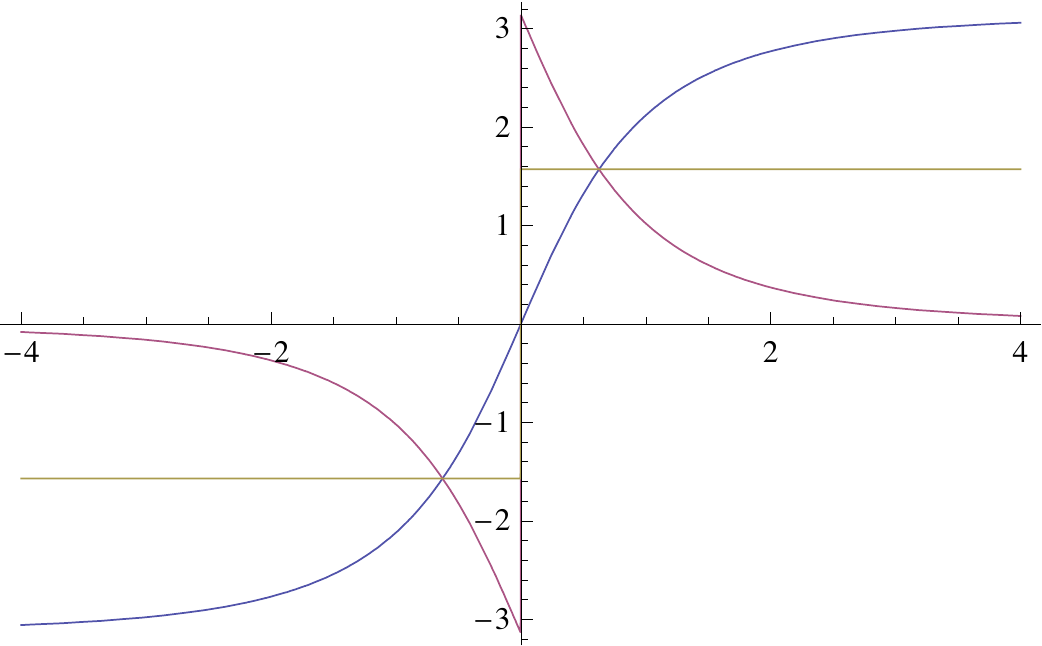}
  \caption{\small  Phases of the cycles $C_{12}$ (blue line), $C_{13}$ (purple line) and $C_{13}$ (grey line) in the generic large $N$ model with widened cuts. The phases are plotted against the distance $\delta$ of cut 2 to the real axis. There are three possible candidates for wall crossing. They correpond to the points where the three phases coincide (including $\delta=0$).}
  \label{cuts}
\end{center}
\end{figure}

To determine the existence of the stability wall, we can study the locus where the BPS phases of the 3-cycles $C_{12}$, $C_{23}$ (and consequently $C_{13}$) align, so that the decay is possible. The phase of the cycle $C_{ij}$ is given by the phase of the integral
\begin{equation}\label{instantonaction}
\int_{C_{ij}}y(x)dx
\end{equation}
which can be evaluated along any path $C_{ij}$ that joins the extreme points $x_i$ and $x_j$. 
Despite the lack of a simple analytic expression for the integral of $y(x)$, we can evaluate numerically the phases of its integral along the different paths. In figure \ref{cuts} we plot the phases of $C_{12}$, $C_{23}$ and $C_{13}$ versus the distance $\delta$ of the cut ${\cal{C}}_2$ to the real axis for a generic saddle point configuration. We can see that in fact there are two points where the three phases align, supporting the required stability wall.

One may worry that the configuration with an eigenvalue distribution in cut 2, which is a maximum, is unstable and may lead to exponentially enhanced instanton amplitudes. 
However the phenomenon persists in other examples where this caveat is absent. For instance, by simply taking the cut ${\cal{C}}_2$ empty and collapsed to a point, the conclusions are unchanged.

A full analytic computation of the BPS phases is possible in a particular limit of the above system, for the matrix model with a single eigenvalue $N=1$, and the three cuts shrunk to three double points. The resolvent function $y(x)$ is just the derivative of the potential $W(x)$:
\begin{equation}
y(x)=(x-x_1)(x-x_2)(x-x_3)
\end{equation}
Although this does not correspond to a geometric regime of the topological string (i.e. large $N$ matrix model), it still maintains the basic properties of its wall crossing.

\begin{figure}
\begin{center}
  \includegraphics[width=10cm]{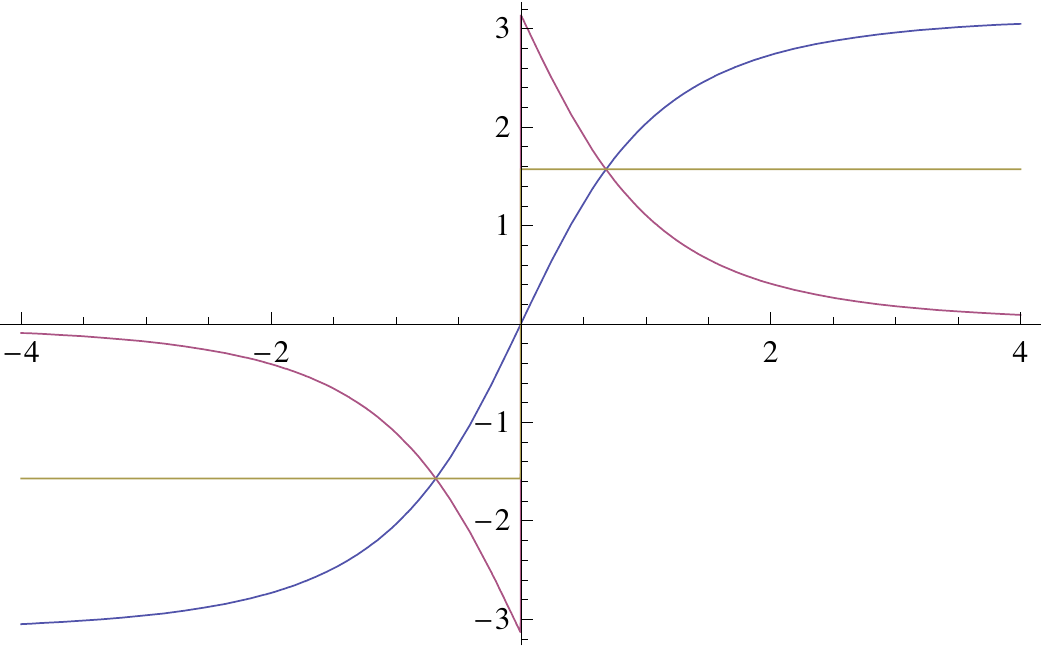}
  \caption{\small Same as figure \ref{cuts}, but for the case where there is just one eigenvalue distributed among the extrema of the potential $W$, and the cuts are collapsed to points. Of the three points where the phases align, only those with $\delta\neq 0$ represent a real line of stability \cite{Townsend:1990}.}
  \label{Townsend}
\end{center}
\end{figure}

The model with cuts collapsed to points can also be regarded as the classical regime of the 4d $\NN=1$ gauge theory in a Dijkgraaf-Vafa duality with the matrix model, in which the strong dynamics scale of the unbroken gauge factors is ignored. The analysis of BPS phases in the matrix model is actually isomorphic to an analysis (in the context of domain walls in Wess-Zumino models) in \cite{Townsend:1990}. The picture of BPS phase in this case is shown in figure \ref{Townsend}, showing the existence of an stability wall.

\section{$\NN=1$ superpotentials and their wall crossing}
\label{nequal1}

In previous sections we have gained a good understanding of threshold wall crossing for D-brane instantons in 4d $\NN=2$ theories. It would be interesting to extend this understanding to 4d $\NN=1$ theories, where such transitions are particularly important, as the only wall crossing phenomenon for D-brane instantons contributing to the superpotential. In this section we describe mechanisms for reduction of supersymmetry, 
and their interplay with the T-duality between particles and instantons. The latter is shown to still underly the continuity of non-perturbative effects across stability walls.
 
\subsection{Introduction of fluxes}
\label{fluxes}

A standard mechanism to reduce the amount of supersymmetry is the introduction of closed string fluxes \cite{Dasgupta:1999ss,Giddings:2001yu,Becker:2001pm} (already exploited for a different purpose in the context of matrix model / gauge theory duality \cite{Dijkgraaf:2002fc,Dijkgraaf:2002vw,Dijkgraaf:2002dh}). In this section we compute non-perturbative superpotentials from D-brane instanton sums in 4d $\NN=1$ flux compactifications, and their threshold wall crossing, by relating them to the underlying 4d $\NN=2$ theories. The analysis ignores the possible presence of other ingredients breaking $\NN=2$ supersymmetry, like orientifold planes, to be discussed in section \ref{oplanes}. 

The effect of fluxes on D-brane instantons has been actively investigated. In particular  fluxes can turn D-brane instantons with 4 fermion zero modes into D-brane instantons with 2 fermion zero modes, allowing them to contribute to the non-perturbative superpotential. This has been established from the coupling of the D-brane instanton fermion zero modes to the flux, computed using D-brane action techniques \cite{Bergshoeff:2005yp,Tripathy:2005hv,Kallosh:2005gs,Park:2005hj}, or CFT correlators
\cite{Billo':2008sp,Billo':2008pg}. Unfortunately these microscopic techniques  must be applied to individual instantons, and cannot take advantage of the powerful resummation techniques of the underlying $\NN=2$ theory. The latter is however fully exploited in the
macroscopic effective field theory technique proposed in  \cite{Uranga:2008nh}. Starting with the effective 4d $\NN=2$ theory of the flux-less  compactification, including non-perturbative effects from resummed D-brane instantons, the effect of fluxes is simply described by the  introduction of the flux superpotential \cite{Gukov:1999ya}. The non-
perturbative superpotential is automatically reproduced by the evaluation of Feynman diagrams involving the spacetime interactions from fluxes and instantons.

More quantitatively, the effective action in $\NN=1$ terms can be written
\beqa
\int \, d^2\theta\, d^2{\ov \theta}\, K(Z,{\ov Z})\, +\, \int \, d^2\theta\, W_{\rm flux} (Z)\,
\eeqa
where $Z$ denotes the $\NN=2$ hypermultiplet moduli written in $\NN=1$ terms, and $K(Z,{\ov Z})$ encodes the moduli space metric. The $\NN=1$ D-term can be written as an F-term as 
\beqa
\int \, d^2\theta\, d^2{\ov \theta}\, K(Z,{\ov Z})\, \simeq\, \int\, d^2\theta\, \frac{\partial^2 K}{\partial{\ov Z}^2}\, {\ov D}{\ov Z}\,{\ov D}{\ov Z}
\eeqa
where ${\ov D}{\ov Z}$ is an anti-chiral multiplet whose lowest component is the 4d fermion in the $Z$. Integrating out the moduli, the non-perturbative superpotential in the flux compactification is \cite{Uranga:2008nh}
\beqa
W_{\rm non-pert.} \, =\,\frac{\partial^2 W_{\rm flux}}{\partial Z^2}\, \, \frac{\partial^2 K}{\partial{\ov Z}^2}\,
\label{fluxdrop}
\eeqa
The net effect is therefore to turn the $\NN=2$ F-term into an $\NN=1$ superpotential term. 

This can be applied to the introduction of fluxes for the 4d $\NN=2$ compactifications in Section \ref{3daction}. For example, consider the concrete example of type IIB on the resolved conifold, or rather the equivalent mirror picture of type IIA on the deformed conifold. The flux configuration we consider is $M$ units of NSNS 3-form flux through the $\IS^3$ $A$-cycle, and $-K$ units through the (suitably regularized) non-compact dual $B$-cycle, plus possibly RR fluxes along non-compact cycles to preserve $\NN=1$ supersymmetry. Using the periods of $\Omega$, the flux superpotential is given by
\beqa
W_{\rm flux}\, \simeq \, \int_{X}\, H_3 \wedge (\, \frac{1}{g_s}\, {\rm Re\, }\Omega+iC_3)\, =\, \frac{1}{2\pi i} M\, Z\, \ln Z \, -\, K\, Z
\eeqa
where $Z=|z|/g_s+ix$, with $z$ the complex structure modulus and $x$ the RR 3-form along $\IS^3$. The model is mirror to the IIB conifold model in \cite{Giddings:2001yu}.
The modulus $Z$ is stabilized at $Z_0=\exp\, (-2\pi \, K/M)$. 

The D2-brane instanton corrections to the moduli space metric were computed in \cite{Ooguri:1996me} (and have been derived directly  from (\ref{vandoreniib}) in \cite{Saueressig:2007dr}). The relevant component for our computation is
\beqa
K_{{\bar Z}{\bar Z}}& =& \sum_{m\neq 0}\,  C_m(z) \, \exp\, \left[\,-2\pi\, \left(\, \frac{|mz|}{g_s}\, -i\, mx \,\right) \,\right]\, \nonumber \\
{\rm with} & & C_m(z) \, =\,   \sum_{n=0}^\infty \, \frac{\Gamma(\frac 12+n)}{2\sqrt{\pi} \, n!\, \Gamma(\frac 12-n)}\, \left(\, \frac{g_s}{4\pi\, |mz|}\,\right)^{n+\frac 12} 
\label{ssov}
 \eeqa
Using (\ref{fluxdrop}), the $\NN=1$ non-perturbative superpotential reads
\beqa
W_{\rm n.p.}\, \simeq \, \frac{M}{Z_0} \,  \sum_{m\neq 0}\,  C_m(z_0) \, e^{-2\pi\, mZ_0} 
\eeqa
where $z_0={\rm Re} Z_0$.
Notice that the moduli in the prefactor should be considered as fixed at their vevs at the minimum, as the above superpotential is valid at scales below the moduli stabilization scale. 

Concerning out interest in threshold wall crossing, the close relation between D-brane instantons in the parent $\NN=2$ model and its $\NN=1$ flux descendant makes it clear that the continuity of the superpotential across threshold walls is automatically encoded in the continuity of the GV invariants in the $\NN=2$ theory.

Remarkably the present setup provides the full non-perturbative superpotential arising from an infinite sum over multiwrapped D2-brane instantons. To our knowledge this is the first time such contributions to the superpotential can be successfully resummed (see \cite{Grimm:2007xm} for partial results in this direction, and
 \cite{Camara:2007dy,Camara:2008zk,Blumenhagen:2008ji} for multi-wrapped instanton contributions to other quantities, motivated by heterotic-type I duality). We hope this kind of result to have an interesting impact on model building applications of D-brane instantons.

\subsection{Introduction of gauge D-branes}
\label{gauge}

We have seen that multiwrapped instanton contributions are quite generic in $\NN=2$ and $\NN=1$ non-perturbative effects. This may seem to lead to a potential puzzle in a different 4d $\NN=1$ context, in which the reduction from $\NN=2$ is obtained by introducing 4d spacefilling D-branes, leading to 4d $\NN=1$ gauge sectors. In such models, certain D-brane instantons (wrapped on the same internal cycles as the gauge D-branes) admit the interpretation of gauge theory instantons. In many examples, the non-perturbative effects in 4d $\NN=1$ gauge theories are ensured, by R-symmetry and holomorphy, to arise only at the 1-instanton level. In these models, such macroscopic considerations forbid the contribution from multiwrapped instantons. In this Section we take a small detour to understand microscopically the absence of multiwrapped instantons, emphasizing the key differences with previous systems where they are present.

We focus on a prototypical example where the non-perturbative effect is an ADS superpotential for $N_f=N_c-1$ SQCD \cite{Affleck:1983mk}. Its derivation from D-brane instanton physics has been discussed e.g. in \cite{Akerblom:2006hx}, however assuming (rather than deriving) the absence of multiwrapped instanton contributions. For our purposes, it will be enough to consider the $N_c=1$ analog of the above ADS superpotential, generated by a D-brane instanton on top of a single 4d gauge D-brane, see \cite{Petersson:2007sc} for a detailed analysis.

For concreteness we consider type IIA on a deformed conifold geometry $X$, with a D6-brane wrapped on $\IS^3$, whose non-perturbative effects can be computed from a T-dual type IIB on $X\times \IS^1$, with a D5-brane on $\IS^3$ and located at a point in $\IS^1$. Namely the sum over D2-brane instantons on $\IS^3$ maps to a one-loop diagram of D3-brane particles running in a loop with momentum on $\IS^1$. As in previous systems, the D3-brane particle momentum maps to the T-dual D2-brane instanton number. We need to show that due to the presence of the D5-brane, the 4d $\NN=1$ superpotential arises only from D3-brane particles with zero momentum on $\IS^1$. This follows from looking at the fermion zero modes on the D3-brane particle worldline, which can be obtained from the open string spectrum. In the D3-D3 open string sector there are four fermion zero modes $\theta^\alpha$, ${\ov \tau}_{\dot\alpha}$, and in the D3-D5 sector there are one complex bosonic mode $b_{\dot\alpha}$ and two fermionic zero modes  $\beta_{\dot\alpha}$, ${\ov \beta}_{\dot\alpha}$. 
D3-brane particles with zero momentum on $\IS^1$ have a quantum wavefunction which is constant on $\IS^1$, and interact with the localized D5-brane. This is reflected by a non-vanishing worldline coupling involving the D3-D5 zero modes of the form
\beqa
S_{\rm D3\, ferm.}\, =\, {\ov \tau}^{\dot\alpha} \,(\, b_{\dot\alpha} {\ov \beta}\, +\, {\ov b}_{\dot\alpha}\, \beta\,)
\eeqa
This lifts all fermion modes except the two fermion zero modes $\theta^\alpha$, and allows the D-branes to contribute to the superpotential. For D3-branes with non-zero momentum, their interaction with the localized D5-brane is weighted by the average of their wavefunctions at the D5-brane location, namely
\beqa
\int dx_0\, e^{ik(x-x_0)} \, =\, 0
\eeqa
The D3-D5  couplings are absent and the D3-branes have too many fermion zero modes to contribute to the superpotential. An identical argument can be run in the more general case of configurations realizing $N_f=N_c-1$ SQCD, with $N_c$ D5-branes on $\IS^3$, and $N_f$ D5-branes on a dual 3-cycle. Hence we recover a result consistent with the expectation from gauge theory instanton physics. 

\subsection{Introduction of orientifold planes}
\label{oplanes}

In this section we consider reducing to 4d $\NN=1$ by an orientifold quotient. The main effect of such orientifolds is that they can project out certain fermion zero modes of D-brane instantons, allowing them to contribute to the superpotential. For instance for D1-brane instantons, this would correspond to wrapping them on holomorphic curves of $RP_2$ topology. Clearly a systematic computation of such 4d $\NN=1$ superpotentials would require a formulation of topological string in orientifold models (dubbed real topological string). Recent progress in this field (see e.g. \cite{Acharya:2002ag,Walcher:2007qp, Krefl:2008sj,Brunner:2008bi,Krefl:2009md}) holds the promise of such applications, although the connection of the real topological strings to the physical orientifold theories is not fully understood. However, the existence of a GV interpretation for real topological string amplitudes almost guarantees the applicability of some $\NN=2$ lessons to $\NN=1$. Here we show that the particular lesson of relating D-brane instantons to D-brane particle loops by T-duality, and threshold bound states to multi-instanton processes, survives in $\NN=1$ orientifold theories, and explains the continuity of non-perturbative $\NN=1$ superpotentials through threshold walls. This should be related to the continuity of unoriented GV invariants for the real topological string. 

In fact, despite the reduced supersymmetry, the discussion is straightforward.
For concreteness we focus on the explicit example of threshold wall crossing for the D-brane instanton contributing to the superpotential in the orientifolded $X_4$ geometry in Section \ref{nonpertacrosswalls}. Following the logic in Section \ref{threshold-walls}, the continuity of the superpotential is ultimately related to the existence of a threshold bound state in the 2-particle quantum mechanics corresponding to the quiver in Figure \ref{quivers}b. The vanishing of the scalar potential is described by the same equations (\ref{modulispace}), and there is similarly a single 2-particle BPS bound state (with only two fermion zero modes, due to the absence of the modes $\tilde\theta_2$ in the quiver quantum mechanics).
As in the $\NN=2$ case, this threshold bound state 'explains' the conspiracies of multi-instanton processes required to reconstruct the non-perturbative superpotential at the wall. It is straightforward to generalize the discussion to many other examples.

\section{More general D-branes charges}
\label{general-charges}

The computation of non-perturbative effects from D-brane instantons with general charges  is an important question, with much recent
progress, see \cite{Alexandrov:2008ds,Alexandrov:2008nk,Alexandrov:2008gh,Alexandrov:2009zh}, yet with open questions.
In this Section we consider the extent to which topological strings, can describe effects from general D-brane instantons, beyond the D1/D(-1) sector. We suggest some interesting connections, supporting the idea that the topological string underlies the continuity of non-perturbative effects across general lines of marginal stability (and thus also the wall crossing formula in \cite{ks}), beyond those in Section \ref{marginal-walls}.

\subsection{D6-brane charge in topological strings}

We have repeatedly mentioned that the topological string only includes effects from  the sector of D2/D0-brane charge. However there is a trick \cite{Gaiotto:2005gf, Dijkgraaf:2006um} that allows to consider more general configurations describing bound states of one D6-brane with induced D2/D0-brane charges, as we review following \cite{Dijkgraaf:2006um}.  

The basic idea is to consider a new kind of state in M-theory on $X\times 
\IS^1\times {\tilde \IS}^1$, given by a KK monopole, described as a Taub-NUT (TN) geometry $K_4$ with $\IR^3$ base filling the (euclidianized) 3d Minkowski directions, and circle fiber along ${\tilde \IS}^1$. In the presence of such object, we have M-theory on $X\times {\IS}^1\times K_4$, where $K_4$ asymptotes to ${\tilde S}^1\times M_3$ but is  topologically $\IR^4$. One considers this geometry, in the presence of a second-quantized gas of spinning particles from M2-branes wrapped on holomorphic cycles of $X$.
The relation comes from two possible type IIA reductions of this configuration. 

$\bullet$ First, by shrinking $\IS^1$, one recovers type IIA theory on $X\times K_4$, which topologically is $X\times \IR^4$. The M2-branes become D2-branes, with momentum along $\IS^1$ becoming D0-brane charge, and with momentum along ${\tilde \IS}^1$ giving angular momentum of this D2/D0-brane particle. The result is a partition function of a second quantized system of D2/D0-brane particles, namely
\beqa
Z_{\rm top.}'\, =\, \exp\, F_{\rm top}'
\label{exp-gv}
\eeqa
where, using the GV interpretation, $F_{\rm top}'$ is the topological string free energy without the contribution from the constant maps. Note that this quantity is determined in terms of the GV invariants.

$\bullet$ Reducing instead along ${\tilde \IS}^1$, one obtains type IIA on $X\times \IS^1$. The TN geometry becomes a D6-brane wrapped on $X\times \IS^1$, bound to a set of D2-branes (from the M2-branes) and D0-branes (from momentum along $\IS^1$)
 \footnote{Inclusion of D4-brane charge can be attempted by using monodromies of the NSNS B-field, at least modulo wall crossing phenomena in this process. Given these subtleties, we focus our discussion on subsets of states with no D4-brane charge.}. 
This yields the partition function of the world-volume theory on a D6-brane wrapped on $X$, bound with D2 on the class ${\bf k}\in H_2(X,\IZ)$ and $q_0$ units of D0 charge, normalized to the partition function of a pure D6-brane with no induced charges, essentially \cite{mnop}
\beqa
Z_{\rm top.}'\, =\, \sum_{{\bf k}, q_0}\, N_{{\bf k},q_0}\, e^{\lambda (\,q_0+t_a k_a\,)}
\label{dt}
\eeqa
The multiplicities $N_{{\bf k},q_0}$ count the BPS groundstates of the D6/D2/D0 system and suffer wall crossing. In a particular chamber in moduli space they are given by the mathematical Donaldson-Thomas (DT) invariants \cite{Donaldson:1996kp}.

The above argument is usually interpreted as relating the topological string on $X$ with the number of BPS groundstates of the D6/D2/D0-theory on $X$. In such terms, the relation can only hold in a particular chamber in moduli space, since the index of BPS D6/D2/D0-particle suffers wall crossing while the GV invariants do not. On this chamber, which we refer to as topological chamber, the equality between (\ref{exp-gv}) and (\ref{dt}) provides a rigorous relation between the mathematical DT and the GV invariants, which we exploit below. Nevertheless, in the following Sections we argue the existence of a (to our knowledge, new) universal relation, valid throughout moduli space, and deeply related to the interpretation of the topological string as computing non-perturbative D-brane instanton effects, which are continuous upon wall crossing.

\subsection{D5/D1/D(-1)-instanton wall crossing and topological strings}

The key observation is that the M-theory 9-11 flip actually relates the GV topological string with a D6/D2/D0-brane on $X\times \IS^1$, namely with a {\em 3d compactification of the system of 4d D6/D2/D0-brane particles}. Moreover the D6-brane worldvolume theory is second-quantized in the argument, suggesting that we actually describe it as a  first-quantized BPS D6/D2/D0-brane particle running along the $\IS^1$ in a one-loop Schwinger diagram. Using our by now familiar c-map, the 9-11 flip of the GV topological string is actually computing non-perturbative D5/D1/D(-1)-brane instanton effects in a T-dual 4d type IIB on $X$. Computations of these effects are discussed in Section \ref{computation}.

This observation has several important (and closely related) implications regarding wall crossing:

$\bullet$ Since D5/D1/D(-1)-brane instanton effects in IIB are computed by the GV topological string (via the 9-11 flip), they are continuous throughout moduli space. Namely are continuous across walls of marginal stability of the underlying D-brane instantons.

$\bullet$ The index of BPS D6/D2/D0-brane 4d particles jumps across walls of marginal stability, but their non-perturbative effects upon $\IS^1$ compactification to 3d must be continuous (being determined by GV invariants). This is possible only due to the mechanism in \cite{Gaiotto:2008cd} and the wall crossing formulas of \cite{ks}. 
The connection between D-brane instanton effects and the topological string connection implies these relations, in that it leads to manifestly continuous non-perturbative effects.

$\bullet$ Beyond the well-known relation between GV invariants and the index of 4d BPS D6/D2/D0-particle states \emph{in the topological chamber} (where the latter are given by DT invariants), we claim that there is a precise relation between GV invariants and the index of 4d BPS D6/D2/D0-particle states \emph{throughout moduli space}. In any given chamber, the 3d non-perturbative effects from the 4d particles on $\IS^1$ must reproduce the GV topological string partition function as described in the 9-11 flip argument. Conversely, the D6/D2/D0 indices in the different chambers could be generated by expanding the GV  partition function in different basis of instantons (corresponding to those descending from the different sets of BPS particles in the 4d lift).

This extended relation is to the best of our knowledge new.
It would be very interesting to work this out explicitly in examples where the DT invariants are known throughout moduli space, e.g. \cite{Jafferis:2008uf,Chuang:2008aw}.

\subsection{Computation of D5/D1/D(-1)-instanton effects}
\label{computation}

In this section we describe more quantitatively the non-perturbative D5/D1/D(-1)-brane instanton corrections to the IIB hypermultiplet metric. The preliminary conclusion is that
the physical interpretation of $Z_{\rm top}'$ in terms of D5/D1/D(-1)-instantons is limited by the present ability to describe quaternionic Kahler metrics, and allows the inclusion of non-perturbative effects in a certain linear approximation. In the topological string chamber, where the BPS multiplicites are the mathematical DT invariants, the linear approximation corresponds to truncation onto single-instanton processes. As one moves to other chambers, however, the linear approximation contains information from certain multi-instanton processes as well.

A second limitation concerns the sector of charge under consideration. The topological string allows to describe arbitrary bound states of D2/D0-branes with one D6-brane, but it does not allow the inclusion of anti-D2 brane states. Therefore the results in this section are restricted to (potentially multi-) instanton processes with total charge corresponding to one D6-brane and to positive D2/D0-brane charge. A more complete description thus remains as an open question.

The discussion of non-perturbative effects from D-brane instantons with mutually non-local charges has been considered in \cite{Alexandrov:2008gh,Alexandrov:2009zh}. Since the instantons break too many isometries of hypermultiplet moduli space, the description of the metric requires a twistor formalism, which provides the basis of our coming discussion. Sketchily (see references for details), the construction of the quaternionic Kahler metric on the hypermultiplet moduli space ${\cal {M}}$ is encoded in a contact structure on its twistor space ${\cal{Z}}$, which is a $\IP^1$ fibration over ${\cal {M}}$. The twistor space can be covered by open sets, which are locally flat when expressed in the so-called Darboux coordinates. The contact structure specifies the set of transition functions (contact transformations) between Darboux coordinates in different patches. All the information is encoded in a set of holomorphic functions, subject to some contraints. 
For cases with enough isometries, one of these functions called the contact potential becomes the $\NN=2$ tensor potential mentioned in section \ref{iibinstantons}.

The perturbative moduli space metric is recovered from a twistor space with three patches. The correction to the metric from brane instantons can be implemented by the introduction of additional patches in the twistor space, with the corresponding contact transformations.  In the linear approximation, where instanton effects are considered a small deformation of the structure of the perturbative theory, the information is encoded in a single holomorphic transition function, which for D-brane instantons with mutually local charges is of the form \cite{Alexandrov:2008gh}
\beqa
G_{\rm local}\, =\, \frac{1}{(2\pi)^2}\, \sum_{k_\Lambda} \, n_{k_\Lambda}\, {\rm Li}_2\, \left( \, 
e^{-2\pi i \, k_\Lambda \xi^\Lambda}
\, \right)
\eeqa
where $k_\Lambda$ denote the instanton integer charges, and $\xi^\Lambda$ are the moduli, so the exponent reproduces the instanton central charge.

A particular case of the above formula has already appeared in \eqref{dilogmembrane}, as the computation of the D2/D0-brane one-loop diagram, or D1/D(-1)-instanton non-perturbative effects. There, $k_0$ is the D1-brane worldvolume flux quantum, other $k_a$ label the 2-cycle homology class, the multiplicities $n_{k_{\Lambda}}$ correspond to the corresponding genus zero GV invariants, and the dilogarithm sums over wrapping numbers of such D1-branes. 

In the present situation, the topological string (via the 9-11 flip) dictates a simple generalization of this computation, namely a one-loop diagram of D6/D2/D0-brane particles. These are simply added in the topological string description, so the result is
\beqa
G_{\rm D5/D1/D(-1)}\, =\, \frac{1}{(2\pi)^2}\,\sum_{{\bf k},q_0}\,  N_{{\bf k}, q_0} \, {\rm Li}_2 \, (e^{\, l^\Lambda\rho_\Lambda \, -\, k_\Lambda \xi^\Lambda})
\label{symplect}
\eeqa
where the central charges of the different D-branes are displayed by introducing a symplectic basis of electric and magnetic charges $(k_\Lambda, l^\Lambda)$, and moduli $(\rho_\Lambda, \xi^\Lambda)$. This is precisely of the form considered in  \cite{Alexandrov:2008gh} for the corrections from D-brane instantons with mutually non-local charges, in the linear approximation mentioned above. 

In the context \cite{Alexandrov:2008gh}, there was no clear interpretation for the D-brane multiplicities. In our setup however, the multiplicities $N_{{\bf k}, q_0}$ have a clear interpretation as the 4d D6/D2/D0-particle indices in the topological chamber, namely the mathematical DT-invariants. Given the derivation of the above result from the 9-11 flip in M-theory, the sum runs over all possible D2/D0-brane states, without inclusion of antibranes (as these are the states that can be bound to a D6-brane in the topological chamber). 

Nevertheless, our statement is that the above expression provides the correct non-perturbative effects even if one moves to other chambers, even if some of these D6/D2/D0-brane states disappear from the BPS spectrum. In order to illustrate the point, consider a situation where a state $\gamma$ disappears through a primitive wall crossing $\gamma\to \gamma_1+\gamma_2$. Restricting to the relevant sector, on the stable side the contribution to (\ref{symplect}) is
\beqa
N_{\gamma_1}\, {\rm Li}_2(e^{-Z_{\gamma_1}})\, +\,  N_{\gamma_2}\, {\rm Li}_2(e^{-Z_{\gamma_2}})\, +\, N_{\gamma}\, {\rm Li}_2(e^{-Z_\gamma})
\label{linear-primitive}
\eeqa
On the other side, the first two terms are still generated by $\gamma_1$ and $\gamma_2$. The last term cannot be generated by $\gamma$ since it is unstable. However it is simple to argue that it is generated by a 2-instanton process involving both $\gamma_1$ and $\gamma_2$. This follows from the considering the linear approximation to the wall crossing formula \cite{ks}, as we explain.

Following the discussion in \cite{Alexandrov:2008gh,Alexandrov:2009zh} (inspired in \cite{Gaiotto:2008cd}), the correction to the hypermultiplet moduli space metric from a D-brane instanton with charge $\gamma$ and BPS phase $\theta_\gamma$ is described as a change in the contact structure of the twistor space ${\cal{Z}}$. Denoting $z$ a complex coordinate on the $\IP^1$ fiber of ${\cal{Z}}$, the instanton introduces new patches, elongated along the real lines (rays) $\ell_{\pm}$ where the phase ${\rm arg}\, z=\pm \theta_\gamma$. The contact structure is encoded in a transition function across $\ell_{\pm}$, which is a symplectomorphism $U_\gamma $
\beqa
U_{\gamma}\, =\, \exp\, {\rm Li}_2 \left(\, e_\gamma\right) \, \to\, \exp\, {\rm Li}_2 \left(\, e^{-2\pi \,Z_\gamma}\right)
\label{linear}
\eeqa
where the arrow indicates how the symplectomorphism is represented as a transition function as the exponential of the dilogarithm in (\ref{symplect}).
For mutually non-local D-brane instanton charges, such symplectomorphisms do not commute, and form an algebra \cite{ks}
\beqa
[e_{\gamma_1},e_{\gamma_2}]\, =\, (-1)^{\langle \gamma_1,\gamma_2\rangle} \, \langle \gamma_1,\gamma_2\rangle\, e_{\gamma_1+\gamma_2}
\eeqa
The contribution from summing over all instantons is associated to taking the products of such symplectomorphisms, ordered according to the value of the BPS phase $\theta_\gamma$ (e.g. counterclockwise around the origin $z=0$). As one moves in moduli space, the BPS phases change and become aligned at walls or marginal stability, beyond which the ordering of the aligned BPS states changes. The continuity of the moduli space metric requires that the spectrum of BPS states jumps, such that
the products over the BPS spectra is unchanged,
\beqa \label{equalproducts}
\prod_{\gamma^+} U_{\gamma^+}^{\Omega^+(\gamma^+)}\, =\, \prod_{\gamma^-} U_{\gamma^-}^{\Omega^-(\gamma^-)}
\eeqa
where the products run through BPS states with charges $\gamma^{\pm}$ on the sides of a wall, and are ordered according to their BPS phase, and $\Omega$ denotes the BPS multiplicities in the corresponding region.

Going back to our primitive crossing example, the wall crossing formula is
\beqa
U_{\gamma_1}^{N_{\gamma_1}}\, U_{\gamma}^{N_{\gamma}}\, U_{\gamma_2}^{N_{\gamma_2}}\, =\, U_{\gamma_2}^{N_{\gamma_2}}\, U_{\gamma_1}^{N_{\gamma_1}}
\eeqa
The linear approximation to the right hand side directly reproduces (\ref{linear-primitive}). The left hand side reproduces it upon use of the Baker-Hausdorff-Campbell formula, and linearization, with
\beqa
N_\gamma\, =\, (-1)^{\langle \gamma_1,\gamma_2\rangle} \, \langle \gamma_1,\gamma_2\rangle\,
\eeqa
namely, the primitive wall crossing formula in \cite{Denef:2007vg}. The last contribution in (\ref{linear-primitive}) thus arises as a linear piece from the combination of the two instantons $\gamma_1$, $\gamma_2$, as announced.

A similar application of the general wall crossing formula implies that
the topological string answer (\ref{symplect}) is robust under wall
crossing.
This nicely dovetails the fact that the result can be unambiguously
expressed in terms of GV invariants (via their connection with DT
invariants). The amusing part is that the way in which different terms in
(\ref{symplect}) are generated changes from chamber to chamber, and in
particular can involve multi-instanton effects. Our point, however, is
that the topological string is clever enough to ignore all
microscopic complications and provide an answer automatically valid
thoroughout moduli space, at least for instanton processes in the charge
sector under consideration. Although the equality \eqref{equalproducts} is
already known, we claim that, at least at the linear level, it is a
consequence of the fact that \eqref{symplect}, (which computes the $U$'s),
is computed by the topological string, which is manifestly robust across
walls, as opposed to just some function of the GV invariants that might
otherwise change in form from chamber to chamber in the moduli space. It would be interesting to
extend this picture to other charge sectors, with multiple D6-branes, or
including anti-D2-brane charges.
\section{Conclusions}
\label{conclusions}

In this work we have discussed the relation between the GV interpretation of the topological string and 4d non-perturbative effects from D-brane instantons. This connection underlies the continuity of non-perturbative effects through walls of BPS stability. The connection is realized via compactification to 3d, and hence fits nicely with the physical interpretation of the wall crossing formula in \cite{ks}. Along the way, we have made contact with the literature on computation of D-brane instantons effects in 4d $\NN=2$ theories, and reproduced some of those results in a computationally powerful way. In addition, we have also discussed mechanisms to reduce supersymmetry to $\NN=1$ and studied its wall crossing. 
Finally, we have suggested a physical interpretation for the recently discussed non-perturbative effects in matrix model instantons.  
 
There are many open directions to be explored in future work. For instance, the systematic inclusion of  all D-brane charges in the topological string, and a more detailed study of the relation with the microscopic wall crossing formulation in terms of symplectomorphism in the twistor space ${\cal{Z}}$.
 
On a more formal line, it would be interesting to understand the relation of universal quantities (in the sense of being insensitive to wall crossing), like the hypermultiplet space metric, with the universal category of holomorphic branes. In different chambers in moduli space, one expects that a localization restricts the computation of the set of BPS states in that region. The original definition in the universal category would however ensure the universality of the result, and consequently encode the microscopic wall crossing properties of the BPS spectrum.

Finally, it would be extremely interesting to develop additional tools to reduce supersymmetry to $\NN=1$ while keeping the fantastic computational power of $\NN=2$. We suspect that the T-dual viewpoint in terms of a one-loop diagram of BPS particles could provide such a powerful framework, and we expect it to lead to new insights into the phenomenological applications of D-brane instantons.

\bigskip

\begin{center}
{\bf Acknowledgements}\\
\end{center}

We thank I. Garcia-Etxebarria, J. Maldacena, C. Mayrhofer, S. Pasquetti, C. Vafa, T. Weigand, D. Westra and E. Witten for useful discussions. A.M.U. thanks 
M. Gonz\'alez  for encouragement and support. This work  has been supported by the Europea Commission under RTN European Programs MRTN-CT-2004-503369,
 MRTN-CT-2004-005105, by the CICYT (Spain) and the Comunidad de Madrid under project HEPHACOS  P-ESP-00346, and in part by the Austrian Research Funds FWF under grant number P19051-N16. 
P.S. aknowledges financial support from Spanish National Research Council (CSIC) through grant JAE-Pre-0800401.

\newpage

\appendix

\end{document}